\documentclass{aa}

\usepackage{graphicx}
\usepackage{txfonts}
\usepackage{booktabs}
\usepackage{amsmath}
\usepackage{url}
\usepackage{enumitem}
\setlist[enumerate]{label=(\roman*)}
\usepackage{lineno}
\usepackage[hidelinks]{hyperref}

\begin{document}

       \title{Solid-state formation of H$_2$CS and CH$_3$SH in molecular clouds:}

       \subtitle{Carbon-initiated reactions with atomic and molecular hydrogen}

\author{
F. Kruczkiewicz\inst{1}
\thanks{Marie Skłodowska-Curie Actions Individual Fellow}
\and
J. C. Santos\inst{1,2}
\thanks{51 Pegasi b Fellow}
\and
G. Fedoseev\inst{3}
\and
S. Ioppolo\inst{4}
\and
C. M. R. Rocha\inst{1,5}
\and
K.-J. Chuang\inst{1}
}

\institute{
Laboratory for Astrophysics, Leiden Observatory, Leiden University,
PO Box 9513, 2300 RA Leiden, The Netherlands\\
\email{kruczkiewicz@strw.leidenuniv.nl}
\and
Center for Astrophysics,
Harvard \& Smithsonian, 60 Garden St., Cambridge, MA 02138, USA
\and
Xinjiang Astronomical Observatory, Chinese Academy of Sciences,
Urumqi 830011, China
\and
Center for Interstellar Catalysis, Department of Physics and Astronomy,
Aarhus University, Aarhus 8000, Denmark
\and
Department of Applied Cryptography \& Quantum Algorithms,
Netherlands Organisation for Applied Scientific Research (TNO),
The Hague, The Netherlands
}

\date{--}

  \abstract
   {Sulfur chemistry serves as a sensitive probe of the chemical evolution of interstellar environments. Despite the detection of several sulfur-bearing molecules in dense clouds, the dominant formation routes of key species such as thioformaldehyde, H$_2$CS, and methanethiol, CH$_3$SH, on icy dust grains are still poorly constrained. Laboratory experiments simulating grain-surface processes are essential to elucidate the low-temperature reaction pathways that link simple precursors like hydrogen sulfide, H$_2$S, to more complex species.}
    {We investigate carbon-initiated surface reactions involving H$_2$S and atomic or molecular hydrogen under dense cloud conditions, focusing on their contribution to the formation of sulfur-bearing species such as H$_2$CS and CH$_3$SH. We focus on identifying the dominant reaction pathways and constraining the thermal desorption behavior of the resulting products.}
   {Using an ultra-high-vacuum setup, we performed low-temperature co-deposition experiments at 10 K involving H$_2$S, ground-state atomic carbon, C($^{3}$P), and H or H$_2$. Reaction products were monitored in situ by Fourier-transform reflection absorption infrared spectroscopy (RAIRS) and confirmed during warm-up by temperature-programmed desorption (TPD) experiments. The experiments were complemented by gas-phase ab initio calculations to assess the reaction pathways leading to CH$_3$SH formation.}
   {The reaction of atomic carbon with H$_2$S at 10~K leads to efficient formation of H$_2$CS, detected both as monomeric H$_2$CS and as H$_2$CS-derived oligomers, (H$_2$CS)$_x$. In the presence of H atoms, H$_2$CS is further processed to CH$_3$SH by successive hydrogenation. When H$_2$ is present instead of H atoms, CH$_3$SH also forms efficiently through an alternative pathway in which the surface reaction C + H$_2$ generates CH$_2$ intermediates that subsequently react with H$_2$S via H-abstraction and radical recombination. Multiple diagnostic infrared bands, together with characteristic TPD fragmentation patterns, provide strong evidence for solid-state H$_2$CS and CH$_3$SH formation at 10~K.}
  {Experimental results show that H$_2$CS and CH$_3$SH can be efficiently formed in ices containing H$_2$S through carbon-initiated solid-state chemistry at 10~K. In particular, the results support two distinct routes toward CH$_3$SH formation: direct H-atom addition reactions to H$_2$CS and CH$_2$-induced H-atom abstraction from H$_2$S ice followed by radical-radical recombination. These findings expand the set of viable solid-state formation pathways for H$_2$CS and CH$_3$SH and place new constraints on astrochemical models of sulfur chemistry in prestellar environments.}


\maketitle
\nolinenumbers

\section{Introduction}
\label{sec:intro}

The detection of molecules in the interstellar medium provides fundamental constraints on the physical and chemical evolution of star- and planet-forming regions. Molecular inventories, abundance ratios, and spatial and thermal distributions encode the history of formation, processing, and transport of matter, making the origin of simple and complex species a central question in astrochemistry \citep[e.g.,][]{Herbst2009ARA&A,oberg_astrochemistry_2021,Cuppen2024ARA&A}. Although most of the more than 350 molecules\footnote{\url{https://cdms.astro.uni-koeln.de/}} identified in interstellar and circumstellar environments have been detected in the gas phase through their rotational spectra, many abundant H-rich species are thought to form efficiently on icy grain surfaces such as H$_2$O, CH$_4$, NH$_3$, and CH$_3$OH \citep{Ioppolo2008ApJ, Dulieu2010A&A, Qasim2020NatAs, Fedoseev2015MNRAS, Watanabe2002ApJ, Fuchs2009A&A}.

Laboratory solid-state experiments under dense molecular cloud conditions have shown that complex organic molecules (COMs) can form via non-energetic atom-addition and abstraction chemistry, without requiring thermal diffusion of heavy radicals. Instead, reactive intermediates generated in close proximity can undergo prompt, non-diffusive radical-radical recombination reactions \citep[e.g.,][]{Fedoseev2015MNRAS2, chuang_h-atom_2016, chuang_formation_2020}. This mechanism has since been incorporated into modern astrochemical models \citep{Jin2020ApJS, Garrod2022ApJ, Borshcheva2025ApJ}. In the dense and cold interiors of prestellar cores, the majority of molecules are frozen out onto grain surfaces, so interstellar ices constitute the main chemical reservoir at this stage \citep[e.g.,][]{Caselli1999ApJ, Bergin2007ARA&A, Caselli2022ApJ}. Consistent with this picture, infrared observations, including recent JWST results, show that these ices are chemically rich, containing both simple molecules and signatures of O-bearing complex species \citep{Boorgert2015ARA&A, Yang2022ApJ, McClure2023NatAs}.

In this context, solid-state sulfur chemistry remains a long-standing puzzle in its chemical evolution and potential carriers.
The cosmic abundance of sulfur is derived from its ionized form, S$^{+}$, in the diffuse interstellar medium, yielding S/H $\sim 1.7\times10^{-5}$ \citep{Esteban2004MNRAS}, close to the solar value of S/H $\sim 1.3\times10^{-5}$ \citep{Asplund2009ARA&A}. However, observations show that in dense clouds only a small fraction of sulfur is found in the gas phase, with depletion factors of two to three orders of magnitude relative to diffuse environments \citep[e.g.,][]{Ruffle1999MNRAS, Jenkins2009ApJ, Vidal2017MNRAS, Laas2019A&A, Fuente2023A&A}. This so-called missing sulfur problem implies the existence of major sulfur reservoirs that are not directly observable.

In the solid state, OCS is the only S-bearing molecule that has been securely identified in interstellar ices \citep{Palumbo1995ApJ, McClure2023NatAs}, with abundances typically of $0.05$--$0.15\%$ relative to H$_2$O ice \citep{Boogert2022ApJ, McClure2023NatAs}, while SO$_2$ has only been tentatively identified at inferred abundances below $0.1\%$ relative to water \citep{McClure2023NatAs}, highlighting how little is known about solid-phase carriers of sulfur. Gaseous ionic or neutral sulfur atoms are expected to accrete and react with abundant and mobile H atoms on grain surfaces at low temperatures. Therefore, H$_2$S is often assumed to be the main sulfur carrier in ices. However, observational constraints of solid H$_2$S toward prestellar and protostellar sources yield only stringent upper limits, with H$_2$S/H$_2$O $\lesssim 0.6$--$0.7\%$ (and as low as $0.13\%$ in some lines of sight), implying that H$_2$S accounts for at most a minor fraction of the solid-phase sulfur budget \citep{Smith1991MNRAS, JimezesEscobar2011A&A, McClure2023NatAs, Slavicinska2025A&A}.

Moreover, laboratory and observational evidence indicates that a significant fraction of H$_2$S does not remain intact, but is rapidly converted to other sulfur reservoirs. In particular, H$_2$S readily undergoes acid–base reactions with NH$_3$ at low temperatures to form ammonium hydrosulfide (NH$_4^+$SH$^-$), a species that may account for up to $\sim$17–18\% of the sulfur budget in dense regions and could be a major carrier of the ubiquitous 6.85~$\mu$m ice feature \citep{Altwegg2022MNRAS, Vitorino2024MNRAS, Slavicinska2025A&A}. Additional sulfur may be locked into more refractory forms, including sulfur minerals and allotropes \citep{Vidal2017MNRAS, Kama2019ApJ, Laas2019A&A, Shingledecker2020ApJ, cazaux_photoprocessing_2022, Carrascosa2024SulphurChains, Herath2025NatCo}, but the detectability of these reservoirs remains highly uncertain.

In contrast to the limited number of sulfur carriers securely identified in interstellar ices, gas-phase observations reveal a diverse sulfur inventory. Sulfur-bearing molecules have been detected in a wide range of environments, from cold cores to hot cores, hot corinos, and Galactic Center molecular clouds, and include both simple species, increasingly complex organosulfur molecules and sulfur-bearing carbon chains \citep{Agundez2014A&A, Cernicharo2021A&A}. Recent surveys have expanded this inventory with sulfur analogs of O-bearing species, including dimethyl sulfide (CH$_3$SCH$_3$), the S analog of dimethyl ether \citep{SanzNovo2025ApJL}, and larger cyclic S-bearing species such as 2,5-cyclohexadien-1-thione \citep{Araki2026NatAs}. These detections emphasize that sulfur can participate in rich organic chemistry even though its main solid-state and gas-phase reservoirs together remain insufficient to close the sulfur budget \citep[e.g.,][]{Drozdovskaya2018MNRAS, RodriguezAlmeida2021ApJL, Agundez2025A&A}.

In the gas phase, sulfur-containing organic molecules such as thioformaldehyde (H$_2$CS) and methanethiol (CH$_3$SH) are central to understanding how sulfur is incorporated into organic chemistry in the interstellar medium. As direct sulfur analogs of H$_2$CO and CH$_3$OH, they have been detected in a wide range of environments, such as prestellar cores, massive hot cores and low-mass protostars \citep[e.g.,][]{Majumdar2016MNRAS,Jorgensen2016A&A, Drozdovskaya2018MNRAS, Vastel2018MNRAS}. Recent studies have further extended these detections to their deuterated counterparts, with HDCS observed in protostellar environments \citep{Drozdovskaya2018MNRAS},  and, more recently, the first interstellar detection of singly deuterated methanethiol (CH$_2$DSH) in IRAS~16293-2422~B \citep{Bunn2025ApJ}.

The formation of H$_2$CS and CH$_3$SH on grain surfaces has frequently been discussed within a successive hydrogenation framework, often by analogy with CO hydrogenation, in which sulfur-bearing precursors (e.g., CS, HCS, and H$_2$CS) are converted through stepwise H-addition reactions to more saturated products \citep{Majumdar2016MNRAS, Muller2016A&A, Vidal2017MNRAS, Lamberts2018A&A, Laas2019A&A, LeGal2019ApJ, DiGenova2026A&A}. Alternative atom/radical-addition routes, such as CH$_2$ + S $\rightarrow$ H$_2$CS and CH$_3$ + SH $\rightarrow$ CH$_3$SH, have also been included in astrochemical networks \citep{Muller2016A&A}. However, direct laboratory evidence for these proposed solid-state pathways remains limited. Recent experiments have reported H$_2$CS and CH$_3$SH formation upon hydrogenation of CS$_2$ \citep{Nguyen2026MNRAS}, and CH$_3$SH and more complex S-bearing molecule formation upon electron irradiation of H$_2$S and CH$_4$ ice mixtures \citep{Herath2026PCCP}. Nevertheless, experiments directly probing hydrogenation networks starting from CS or H$_2$CS are still missing, in part because work with these reactive and/or hazardous sulfur-bearing precursors is technically challenging. As a result, the feasibility and product outcomes of these hydrogenation pathways under astrophysically relevant conditions remain experimentally poorly constrained.

Alternative formation routes involving the expected first-generation of S-bearing ice, like H$_2$S, have also been proposed to contribute to H$_2$CS formation. \cite{Laas2019A&A} proposed that atomic carbon can react with H$_2$S ice forming HCS which is efficiently converted into H$_2$CS through hydrogenation \citep[]{Laas2019A&A}. This framework shifts the chemical entry point from prerequisite CS to carbon-initiated reactivity in ices involving H$_2$S, which has also been experimentally studied in the gas-phase \citep[]{Kaiser1997JChPh,Deeyamulla2006}.

This carbon-initiated H$_2$S chemistry also motivates considering the role of the hydrogen reservoir itself in the ice reactivity \citep{Chuang2018A&A, Martin-Domenech2024ApJ}. While atomic H is well established as a key driver of grain-surface chemistry, particularly through successive hydrogenation and H-abstraction reactions \citep[e.g.,][]{Watanabe2002ApJ, Fuchs2009A&A, Chuang2016MNRAS, Qasim2020NatAs, fedoseev_hydrogenation_2022, Cuppen2024ARA&A}, the potential contribution of molecular hydrogen (H$_2$) to solid-state reaction networks remains much less constrained. Under appropriate conditions, H$_2$ can become chemically active when interacting with atomic carbon \citep{Lamberts2022ApJ, Ferrero2024ApJ}. In particular, insertion of atomic carbon in H$_2$ has been proposed as an entry point to hydrocarbon chemistry in ices leading to CH$_4$ formation \citep{Lamberts2022ApJ}, and recent astrochemical models suggest that C + H$_2$ reactions in interstellar ices can generate CH$_2$ and CH$_3$ radicals, thus opening new pathways toward complex organic molecules \citep{Garrod2022ApJ, Borshcheva2025ApJ, Krasnokutski_2020}. Solid-state reactions following C + H$_2$ $\rightarrow$ CH$_2$ routes have been proposed as an efficient means to enrich the CH$_x$ radicals in ices, yet they remain untested under laboratory ice-analog conditions. Their efficiency and role in driving subsequent ice chemistry therefore remain poorly constrained.

In this work, we experimentally investigate carbon-initiated solid-state routes linking H$_2$S-containing ices to the formation of H$_2$CS and CH$_3$SH. Using cryogenic ultra-high-vacuum co-deposition experiments, we (i) examine whether atomic carbon reacts directly with H$_2$S to form H$_2$CS (C + H$_2$S $\rightarrow$ H$_2$CS), (ii) test successive H-atom addition to H$_2$CS as a route toward CH$_3$SH (H$_2$CS + 2H $\rightarrow$ CH$_3$SH), and (iii) probe an H$_2$-enabled pathway in which C + H$_2$ chemistry generates the CH$_2$ intermediate which can subsequently react with H$_2$S to form CH$_3$SH. To support the interpretation of CH$_3$SH formation in the experiments involving H$_2$, we complement the laboratory measurements with gas-phase ab initio calculations. These calculations compare the singlet and triplet CH$_2$ + H$_2$S reaction channels and the alternative H$_2$ + H$_2$CS pathway, assessing their energetic feasibility and helping to constrain the mechanism underlying the experimental results. By probing these reaction channels under controlled interstellar ice analog conditions, this work assesses whether atomic carbon can drive the conversion of simple sulfur-bearing ices into more complex organosulfur species under dense-cloud conditions.

\section{Experimental methods}

Experiments were carried out using the SURFRESIDE$^3$ ultra-high vacuum (UHV) setup. This setup has been optimized to experimentally study the solid-state atomic reactions on dust grains under dense cloud conditions and enables surface reactions in interstellar-ice analogs at cryogenic temperatures on a gold-plated substrate, while monitoring solid-state chemistry with Fourier-transform reflection absorption infrared spectroscopy (FT-RAIRS) and gas-phase species during temperature-programmed desorption using a quadrupole mass spectrometer (TPD-QMS). SURFRESIDE$^3$ has been described in detail by \cite{Ioppolo2013RScI} and \cite{Qasim2020RScI}.

Interstellar ice analogs were grown on a gold-plated copper substrate mounted to the tip of a closed-cycle He cryostat at the center of the main chamber that is kept at a base pressure of $\sim 5 \times 10^{-10}$ mbar. The substrate temperature can be controlled between 8 and 450~K by resistive heating and it is read by two Si-diode sensors that hold an absolute accuracy of $\sim 0.5$~K. Co-deposition through independent inlet lines allowed us to combine ground-state C atoms (i.e., C($^{3}$P)), H atoms, H$_2$S, and H$_2$. H$_2$S was admitted through an all-metal leak valve. H atoms were supplied by a hydrogen atom beam source (HABS) using thermal cracking to produce the atom beam. The mixed beam (H + residual H$_2$) was cooled by a nose-shaped quartz tube prior to the surface.
For the H$_2$ experiments, the same line and leak-valve setting were used with the dissociation filament off. The unchanged inlet setting maintains the nominal hydrogen supply to the source. For reporting purposes, Table~\ref{tab:exp_2025} lists a nominal value of $0.5 \times 10^{13}$~molecules~cm$^{-2}$~s$^{-1}$, taken as half the calibrated H-atom flux under the idealized assumptions of complete dissociation in H-atom mode and equivalent beam transmission.

A customized SUKO-A 40 C-atom source based on a commercial design (Dr. Eberl, MBE; \citealt{Krasnokutski2014ApPhL,Albar2017NatSR}) produced a beam dominated by C($^3$P) atoms, with a cluster contamination below $1~\%$. The contribution of electronically excited C($^1D$) is expected to be negligible because its excitation energy is substantially higher than the operating temperature of the source \citep{Qasim2020RScI, Lamberts2022ApJ}. The beam was collimated onto the substrate by a set of apertures to limit the deposition of C atoms on the walls of the UHV chamber. CO and CO$_2$ are known contaminants in the C-beam and their deposition rates were estimated to be of the order of $ 1~-~3~\times~10^{12}$ cm$^{-2}$ s$^{-1}$ and $ 2~-~5~\times~10^{9}$ cm$^{-2}$ s$^{-1}$. The CO deposition rate was lower than that of H$_2$S and H/H$_2$, which is expected to favor encounters of incoming C atoms with these species over CO in a well-mixed ice. Comparable CO and CO$_2$ contamination levels were observed across the experiments, which were performed sequentially under similar conditions.
The contribution of CO contamination is limited due to the non-detection of CO hydrogenation products, such as H$_2$CO and CH$_3$OH. Each atom-beam line (HABS and C-source) was differentially pumped to a base pressure of $\leq 1 \times 10^{-9}$ mbar and connected to the main UHV chamber through a shutter valve.

The mixed ices were deposited at 10 K and monitored in situ using FT-RAIRS with a 700-4000 cm$^{-1}$ range and 1 cm$^{-1}$ resolution. After deposition, the sample was heated at 5 K min$^{-1}$ for TPD. The desorbing species were ionized at 70 eV and detected by a QMS, while the solid phase was tracked simultaneously by FT-RAIRS.

The gaseous samples used in this work were hydrogen sulfide (H$_2$S, Linde Gas, $\geq$99.5\%) and molecular hydrogen (H$_2$, Linde Gas, $\geq$99.999\%).
Carbon filaments for the SUKO-A source were purchased from MBE Komponenten GmbH.

\subsection{Experimental analysis}

The absolute column density $N$ was derived using the Beer--Lambert relation:

\begin{equation}
N{_X} = \frac{ln 10}{A'(X)} \int Abs({\nu}) \, d{\nu},
\end{equation}

\noindent where $Abs ({\nu})$ is the integrated absorbance of IR features and $A'$ is the apparent band strength of the transition. For H$_2$S, we used a setup-specific band strength of $A' = (4.7 \pm 0.1) \times 10^{-17}\,\mathrm{cm\,molecule^{-1}}$ for the applied RAIRS configuration, as determined by \cite{Santos2023A&A}.
The H flux has been calibrated in situ for various H$_2$ flow rates on SURFRESIDE$^3$ and determined to be around 1 $\times 10^{13}$ for the applied experimental conditions. The flux of carbon atoms was determined prior co-deposition experiments when a new carbon filament was installed, following the method described by \cite{Qasim2020RScI}. Briefly, C atoms were co-deposited with a large excess of $^{18}$O$_2$; the reaction between C and $^{18}$O$_2$ yields C$^{18}$O and $^{18}$O, with the latter reacting further with excess $^{18}$O$_2$ to form $^{18}$O$_3$.
The $^{18}$O$_3$ column density, derived from its RAIRS absorption band, was then used as an indirect measure of the C-atom flux reaching the substrate.
All co-depositions were performed with comparable fluxes and deposition times, enabling qualitative comparison of relative band intensities across experiments.

The relative abundances of CH$_3$SH and H$_2$CS were derived from the TPD-QMS data using the following method. For CH$_3$SH and H$_2$CS, molecular ions were used at $m/z$ = 48 and 46, respectively, for quantification. The relative abundance ratio was obtained with the method described by \cite{martin-domenech_uv_2015} using the ratio of the integrated TPD peak areas, \(A(m/z)\), corrected for the electron-impact ionization cross sections, \(\sigma^+\), the fragmentation factor, defined as the fraction of molecules that yield a fragment of mass \(m\) in the mass spectrometer, \(F_F(m)\), and the QMS sensitivity at the corresponding mass channels, \(S(m/z)\):

\begin{equation}
\frac{N_{\mathrm{X}}}{N_{\mathrm{Y}}} =
\frac{A(m/z(X))}{A(m/z(Y))}
\cdot
\frac{\sigma^+(\mathrm{Y})}{\sigma^+(\mathrm{X})}
\cdot
\frac{F_F(Y)}{F_F(X)}
\cdot
\frac{S(m/z(Y))}{S(m/z(X))}
\label{eq:CH3SH_H2CS_ratio}
\end{equation}

\noindent Following \citet{martin-domenech_uv_2015}, equal effective pumping speeds are assumed, so no additional correction for relative pumping speeds is applied in Eq.~\ref{eq:CH3SH_H2CS_ratio}. Since experimental ionization cross sections at 70~eV were not available for these species, \(\sigma^+\) was estimated from the molecular polarizability \(\alpha\) using the empirical relation \(\sigma^+_{\max}(X)=c\,\alpha(X)\), with \(c = 1.48\)~\AA$^{-1}$ \citep{Hudson2003_PCCP_alcohols, Bull2008_IJMS_polarizability}. Adopting \(\alpha = 5.186\)~\AA$^3$ for CH$_3$SH and \(\alpha = 4.218\)~\AA$^3$ for H$_2$CS yields \(\sigma^+(\mathrm{CH_3SH}) = 7.68\)~\AA$^2$ and \(\sigma^+(\mathrm{H_2CS}) = 6.24\)~\AA$^2$. Fragmentation fractions were derived from available mass spectra databases, using NIST for CH$_3$SH and SpectraBase for H$_2$CS. The parameters adopted for all species are summarized in Table~\ref{tab:qms_parameters}.

\begin{table}[h]
\caption{Parameters used to derive the relative abundances of CH$_3$SH and H$_2$CS from the TPD-QMS data.}
\label{tab:qms_parameters}
\centering
\begin{tabular}{lccccc}
\hline
Species & $m/z$ & $\alpha$ (\AA$^3$) & $\sigma^+$ (\AA$^2$) & $F_F(m)$ & $S(m/z)$$^{a}$ \\
\hline
CH$_3$SH  & 48 & 5.186 & 7.68 & 0.33 & 0.18 \\
H$_2$CS   & 46 & 4.218 & 6.24 & 0.39 & 0.17 \\
\hline
\end{tabular}

\vspace{1mm}
\raggedright
\footnotesize
$^{a}$ QMS sensitivity factors adopted from \citet{Chuang2018PhD}.
\end{table}

An overview of the experiments performed is given in Table~\ref{tab:exp_2025} including the deposition temperature, the atomic and molecular fluxes, and the deposition time.

\begin{table*}[t]
\centering
\caption{Overview of the experiments performed in this work}
\label{tab:exp_2025}
\small

\begin{tabular}{llccccccc}
\hline
\# &
  Experiment &
  $T$ &
  H$_2$S &
  C &
  H &
  H$_2$ &
  H$_2$S : C &
  Time \\
 &
  &
  (K)
  &
  (cm$^{-2}$\,s$^{-1}$) &
  (cm$^{-2}$\,s$^{-1}$) &
  (cm$^{-2}$\,s$^{-1}$) &
  (cm$^{-2}$\,s$^{-1}$) &
  : H$_2$/H &
  (s)
  \\ \hline

1 &
  H$_2$S &
  10 &
  $\sim 1 \times 10^{13}$ &
  \dots &
  \dots &
  \dots &
  20 : -- : -- &
  1800 \\

2 &
  H$_2$S + C &
  10 &
  $\sim 1 \times 10^{13}$ &
  $\sim 5 \times 10^{11}$ &
  \dots &
  \dots &
  20 : 1 : -- &
  3600 \\

3 &
  H$_2$S + C + H &
  10 &
  $\sim 1 \times 10^{13}$ &
  $\sim 5 \times 10^{11}$ &
  $\sim 1 \times 10^{13}$ &
  \dots &
  20 : 1 : 20 &
  3600 \\

4 &
  H$_2$S + C + H$_2$ &
  10 &
  $\sim 1 \times 10^{13}$ &
  $\sim 5 \times 10^{11}$ &
  \dots &
   $\sim 0.5 \times 10^{13}$ $^{a}$ &
  20 : 1 : 10 &
  3600 \\

  5 &
  CH$_3$SH $^{b}$  &
  10 &
   \dots &
   \dots &
  \dots &
   \dots &
  \dots &
  3600 \\

\hline
\end{tabular}

\vspace{2mm}

\raggedright

$^{a}$ Nominal H$_2$ flux, taken as half the calibrated H-atom flux assuming complete dissociation (see text).

$^{b}$ For Exp.~5, a total of 20~ML of pure CH$_3$SH was deposited over 3600~s.

\end{table*}

\subsection{Ab initio calculations}

All electronic structure calculations employed the \texttt{MOLPRO} quantum chemistry package~\citep{MOLPRO}. The stationary structures on both singlet and triplet potential energy surfaces (PESs) of CH$_3$SH were optimized at CCSD(T)-F12b/VTZ-F12 level of theory~\citep{KNI09:054104,FEL013:084110}. Harmonic vibrational analyses were subsequently done to confirm the nature of the optimized molecular structures as either minima (all real frequencies) or transition states (only one imaginary frequency); these calculations also yielded harmonic zero-point energies (ZPEs).

To understand the connections between the various transition states and associated minima,
we performed one-dimensional PES cuts along suitable optimized reaction coordinate (ORC) paths~\citep{VAR018:20170145}; this involves the selection of a single reactive coordinate responsible for the envisaged chemical process, with all other
degrees of freedom being fully optimized at every selected discrete (fixed) value of such a coordinate (see, e.g.,~\citet{VAR018:20170145}).
Due to convergence difficulties of the Hartree-Fock (HF) and CCSD(T) wave functions at long-range regions of the PES, as well as the increasing multireference character of the wavefunction, ORC paths involving transition states connecting valence structures to dissociation channels were done at the complete active space self-consistent field (CASSCF) level of theory~\citep{SZA012:108} using the aug-cc-pV$T$Z (AV$Q$Z) basis set~\citep{DUN89:1007,KEN92:6796}. The CASSCF active space consists of 14 electrons in 12 orbitals.

Final structures and energetics were all obtained at the CCSD(T)-F12b/VTZ-F12. Note that, for dissociation channels, asymptotic energies were computed as the sum of the energies of the isolated fragments at infinite separation, each evaluated at the CCSD(T)-F12b/VTZ-F1 level.

\section{Results}

\subsection{\texorpdfstring{Formation of H$_2$CS and CH$_3$SH}{Formation of H2CS and CH3SH}}

\begin{figure*}
    \centering
    \includegraphics[width=\textwidth]{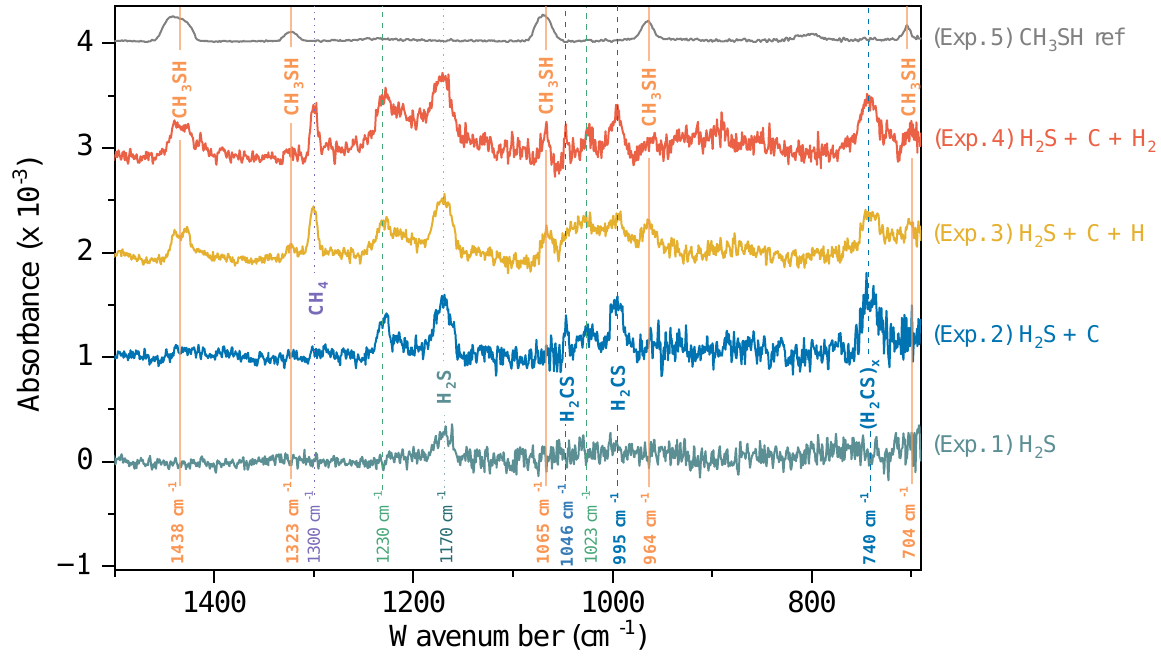}
    \caption{
    RAIRS spectra of the co-deposition experiments performed at 10~K for:
    (Exp. 1) H$_2$S,
    (Exp. 2) H$_2$S co-deposited with atomic C,
    (Exp. 3) H$_2$S + C + H,
    (Exp. 4) H$_2$S + C + H$_2$,
    and (Exp. 5) CH$_3$SH reference, scaled by a factor of 0.1.
    Orange lines indicate the vibrational modes of CH$_3$SH.
    Dashed vertical lines highlight the new absorption features that appear in the H$_2$S + C experiments.
    Dotted lines mark characteristic H$_2$S and CH$_4$ bands present in this spectral region.}
    \label{fig:ftir_1}
\end{figure*}

To directly address the reaction pathways introduced in Section~\ref{sec:intro}, the RAIRS results are organized around three comparisons: the reaction of C atoms with H$_2$S to form H$_2$CS, the further hydrogenation promoted by additional H atoms toward CH$_3$SH, and the formation of CH$_3$SH through a CH$_2$-initiated pathway enabled by H$_2$ co-deposition.
Figure~\ref{fig:ftir_1} shows the IR spectra at 10 K for pure H$_2$S, H$_2$S + C (Exp. 2), and H$_2$S + C + H (Exp. 3) experiments in the spectral range of 1500–690 cm$^{-1}$. For comparison, a reference spectrum of pure CH$_3$SH at 10 K is shown at the top.

\paragraph{(i) H$_2$CS formation in the C + H$_2$S experiment.}

In the spectral window of 1500–690~cm$^{-1}$, the IR spectrum of pure H$_2$S (Exp. 1) is dominated by the $\nu_2$ bending mode at 1170~cm$^{-1}$ \citep{Hudson2018ApJ}. Upon co-deposition of H$_2$S with atomic carbon (Exp. 2), this feature remains present as a parent remnant, while several additional absorption bands appear, indicating the solid-state product formation at 10 K.

In Exp. 2, the feature at 995~cm$^{-1}$ is assigned primarily to the CH$_2$ wagging mode ($\nu_4$) of monomeric H$_2$CS, with a possible contribution from the nearly degenerate CH$_2$ rocking mode ($\nu_6$), in agreement with previous matrix-isolation and solid-state studies reporting this vibration at 987~cm$^{-1}$\citep{Jacox1975JMoSp, Suzuki2007VibSp} and near 990~cm$^{-1}$ \citep{Shao2023JPCL}. The C=S stretching mode ($\nu_3$) of H$_2$CS is intrinsically much weaker; it is observed in this work at around 1046~cm$^{-1}$, compared to 1055.2~cm$^{-1}$ reported by \cite{Suzuki2007VibSp}, a shift of 9~cm$^{-1}$ attributable to matrix and ice-environment effects. This assignment is also consistent with recent high-level anharmonic calculations showing that the $\nu_4$ mode near 1000~cm$^{-1}$ is the strongest infrared feature of H$_2$CS, whereas the $\nu_3$ near 1060~cm$^{-1}$ is significantly weaker \citep{Guo2023RJPCA}.

Furthermore, the formation of H$_2$CS is accompanied by the simultaneous detection of oligomeric forms of H$_2$CS, (H$_2$CS)$_x$, at 740~cm$^{-1}$. This assignment is consistent with the matrix-isolation study of \cite{Jacox1975JMoSp}, who reported the strongest infrared absorption of the cyclic H$_2$CS trimer, s-trithiane, (H$_2$CS)$_3$, at 735~cm$^{-1}$. However, because other H$_2$CS oligomers may also absorb in this spectral region, the present feature cannot be attributed uniquely to s-trithiane. We therefore interpret the 740~cm$^{-1}$ band as evidence for H$_2$CS oligomerization, potentially including contributions from the cyclic trimer and other (H$_2$CS)$_x$ species. Additional weaker bands are observed at 1023 and 1230~cm$^{-1}$ in our experiments, although their exact origins remain uncertain. These features may tentatively arise from minor vibrational modes of H$_2$CS-derived oligomers or clusters.

The detection of (H$_2$CS)$_x$ indicates that newly formed H$_2$CS monomers can undergo association or oligomerization under the present deposition conditions. This assignment is consistent with the propensity of H$_2$CS to form oligomeric or clustered species, whereas isolated monomeric H$_2$CS is typically generated in situ. (H$_2$CS)$_3$ is one of the stable oligomeric forms of H$_2$CS and has been used as a commercially available precursor for H$_2$CS generation via thermal decomposition at 700--800$^\circ$C. Alternatively, monomeric H$_2$CS can be generated by UV photolysis or thermal degradation of suitable precursors such as CH$_3$SH, followed by rapid trapping at low temperature \citep{Jacox1975JMoSp, Suzuki2007VibSp, Muller2019A&A}.

\paragraph{(ii) CH$_3$SH formation in the presence of atomic H.}

In the H$_2$S + C + H experiment (Exp. 3), several new absorption features appear exclusively when H atoms participate in the solid-state reactions, indicating the formation of hydrogenated products. For example, a band attributed to CH$_4$ is observed near 1300~cm$^{-1}$ resulting from the direct hydrogenation of C atoms \citep{Qasim2020NatAs}.

In Exp. 3, the additional IR bands at 964, 1065, and 1438~cm$^{-1}$ are assigned to the CH$_3$ rocking mode ($\nu_{11}$), the CH$_3$ deformation/rocking mode ($\nu_6$), and the CH$_3$ antisymmetric deformation modes ($\nu_4/\nu_{10}$) of CH$_3$SH (see the IR reference spectrum of Exp.~5). Other weaker IR transitions of CH$_3$SH at $\sim$704~cm$^{-1}$ ($\nu_9$), $\sim$800~cm$^{-1}$ ($\nu_7$), and $\sim$1323~cm$^{-1}$ ($\nu_5$) bands, are barely detected due to the relatively high S/N ratio required to identify these weaker features.
The observed band positions and relative prominence are consistent with the laboratory spectra reported of amorphous CH$_3$SH ice at 17~K \citep{Hudson2016PCCP}.

\paragraph{(iii) CH$_3$SH formation in the presence of H$_2$.}

In the IR spectrum of H$_2$S + C + H$_2$ (Exp. 4), the formation of H$_2$CS is again confirmed by the characteristic absorption bands at 995 and 740~cm$^{-1}$, assigned to monomeric H$_2$CS and oligomeric (H$_2$CS)$_x$, respectively. The weaker features at 1046, 1024, and 1230~cm$^{-1}$, also observed in the C + H$_2$S experiment, are present as well. In addition, the CH$_4$ band at 1300~cm$^{-1}$ is detected, consistent with the additional reaction pathway initiated by C + H$_2$ $\rightarrow$ CH$_2$, followed by CH$_2$ + H$_2$ $\rightarrow$ CH$_4$, as reported by \citet{Lamberts2022ApJ}. Besides these products, CH$_3$SH is also clearly identified. The strongest diagnostic CH$_3$SH bands are observed at 964, 1065, and 1438~cm$^{-1}$, while the weaker features at 704 and 1323~cm$^{-1}$ are less prominent, consistent with the relative band intensities observed in pure CH$_3$SH ice (Exp.~5). The positions of the bands and the overall spectral profile closely match those of the reference CH$_3$SH infrared spectrum.

\subsection{Temperature-programmed desorption of products}
\label{sec:TPD}

\begin{figure*}
    \centering
    \includegraphics[width=\textwidth]{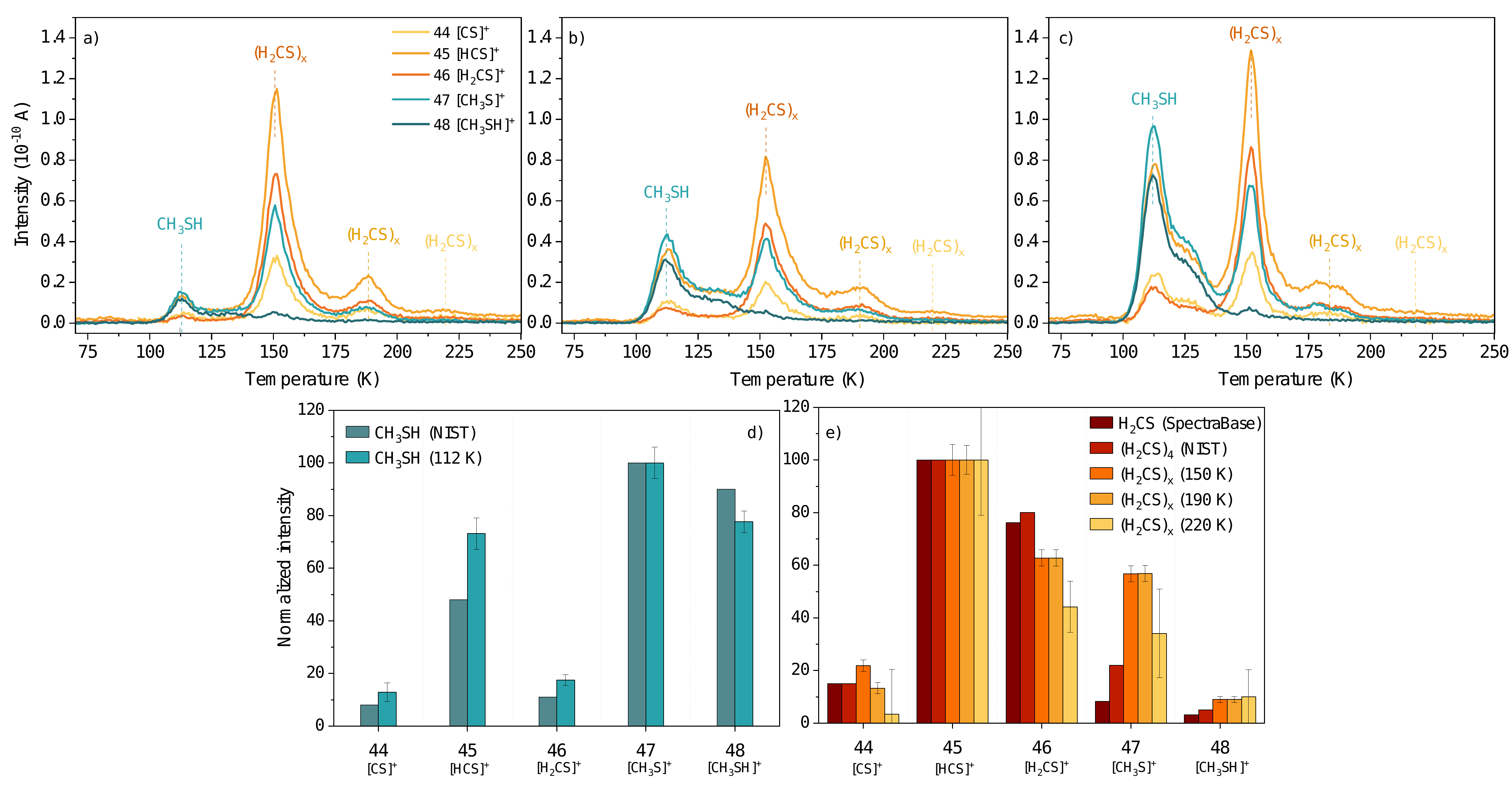}
    \caption{
    Top: TPD mass spectra for ices co-deposited at 10~K in (a) H$_2$S + C (Exp.~2), (b) H$_2$S + C + H$_2$ (Exp.~4), and (c) H$_2$S + C + H (Exp.~3).
    Bottom: comparison of the integrated TPD peak intensities within four temperature intervals (112~K, 150~K, 190~K and 220~K) with the expected fragmentation patterns of CH$_3$SH (d) and H$_2$CS (e) from the literature. Error bars represent the uncertainties arising from the integration fits.
    }
    \label{fig:tpd_cod}
\end{figure*}

After co-deposition, the ice samples were warmed up to 300 K with a temperature ramp of 5 K.min$^{-1}$ to monitor the desorption behavior of the newly formed sulfur-bearing species. The top panels of Fig.~\ref{fig:tpd_cod} show the TPD-QMS data obtained for the (a) H$_2$S + C, (b) H$_2$S + C + H$_2$, and (c) H$_2$S + C + H experiments (Exp.~2,~4 and~3). The bottom panels display the corresponding integrated intensities of the selected mass fragments, compared to the reference mass-fragmentation patterns of (d) CH$_3$SH and (e) H$_2$CS from the NIST\footnote{NIST Chemistry WebBook, \url{https://webbook.nist.gov/}} and SpectraBase\footnote{SpectraBase, \url{https://spectrabase.com}} databases. The integrated intensities shown in panel (d) were obtained from Exp.~3 for the CH$_3$SH desorption feature over 90--130~K, while those in panel (e) were obtained from Exp.~2 for the H$_2$CS-derived oligomer desorption features over 130--170~K, 170--205~K, and 205--240~K, corresponding to the peaks at $\sim$150, 190, and 220~K, respectively.

The first TPD-QMS feature, centered at $\sim$110~K, is attributed to CH$_3$SH desorption. The CH$_3$SH desorption temperature is in agreement with the value reported by \citet{Narayanan2025ApJ} for thin multilayer $^{13}$CH$_3$SH ices. Furthermore, the relative intensities of the selected mass signals (i.e., $m/z$ 44--48) are consistent with the standard CH$_3$SH fragmentation pattern, as shown in Fig.~\ref{fig:tpd_cod}d.

In addition to the CH$_3$SH peak, three TPD-QMS features are detected centered at 150, 190, and 220~K. All three peaks consist of highly similar mass-fragmentation patterns (Fig.~\ref{fig:tpd_cod}e), indicating that they originate from oligomeric species rather than from structurally distinct molecules. For lower mass channels ($m/z$= 44–46), the observed fragmentation pattern closely resembles that of H$_2$CS reported in the SpectraBase database and for the cyclic (H$_2$CS)$_4$ species in the NIST database. In mass spectral databases, H$_2$CS (i.e., SpectraBase) and (H$_2$CS)$_4$ (i.e., NIST) exhibit nearly identical fragmentation patterns. Although these fragmentation patterns do not allow individual oligomers to be identified, their similarity supports the contribution of different H$_2$CS-derived species, collectively denoted as (H$_2$CS)$_x$, to the higher-temperature desorption features.

These high-temperature desorption features occur well above the CH$_3$SH desorption temperature under the same experimental conditions, suggesting that they correspond to species that are more strongly retained in the ice.

\begin{figure}
    \centering    \includegraphics[width=\columnwidth]{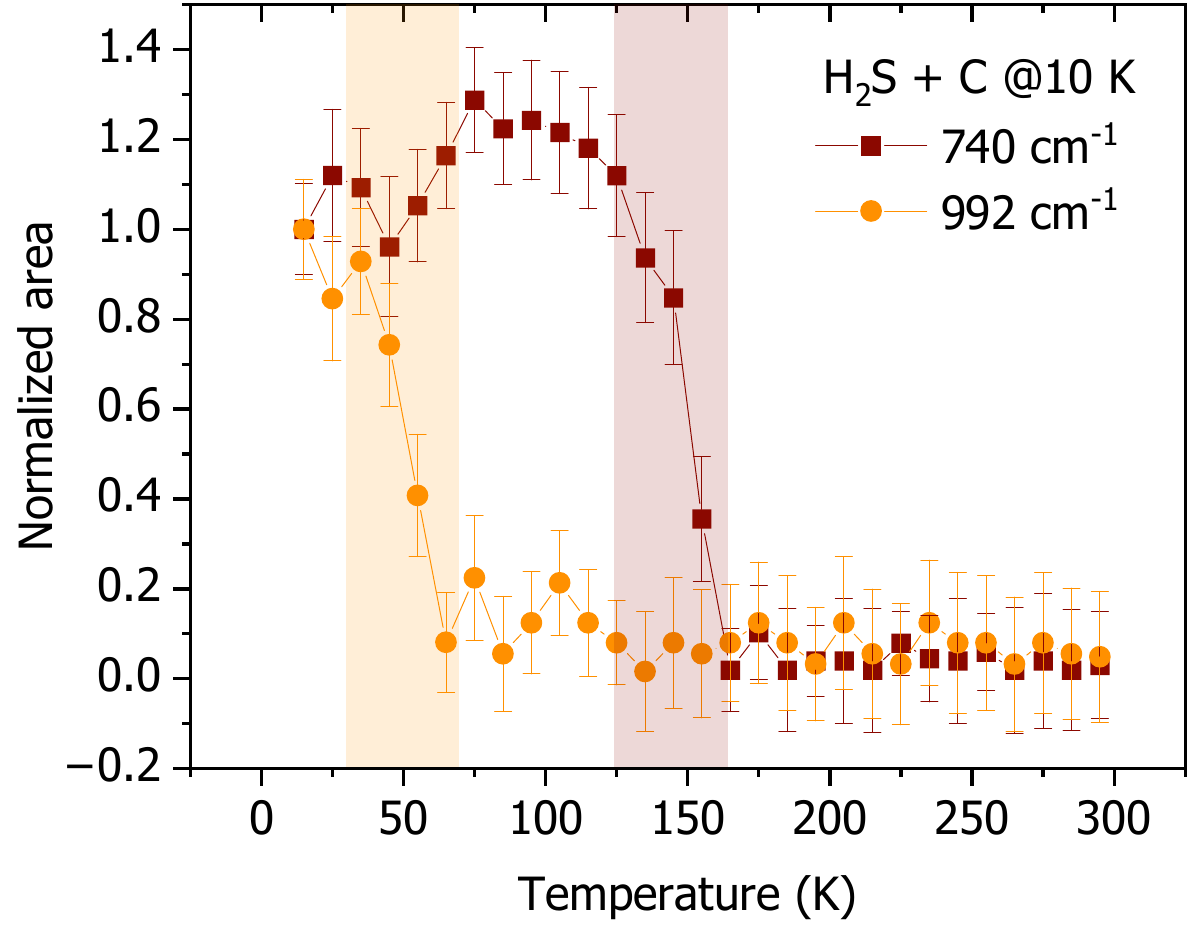}
    \caption{
    Integrated absorbance of the H$_2$CS monomer band at 992~cm$^{-1}$ (circle) and the (H$_2$CS)$_x$ band at 740~cm$^{-1}$ (square) as a function of the temperature during TPD for ices deposited at 10~K, each normalized to its own initial integrated area. Error bars reflect band-integration uncertainties. Connecting lines are for visual guidance only.}
    \label{fig:RAIRS_TPD}
\end{figure}

IR measurements during the TPD experiment provide complementary information on the temperature-dependent behavior of monomeric and oligomeric (H$_2$CS)$_x$. Figure~\ref{fig:RAIRS_TPD} shows the integrated absorbance bands of the H$_2$CS monomer band at 992 cm$^{-1}$ and the oligomeric band at 740 cm$^{-1}$ as a function of temperature during the TPD experiment.

The monomer band decreases rapidly between 30 and 80 K. In the same temperature range, weak $m/z$ = 45 and 46 signals are observed in the QMS data, indicating that a fraction of monomeric H$_2$CS desorbs between $\sim$70 and 95~K (Fig.~\ref{fig:h2cs-monomer-desorption} in Appendix~\ref{app:h2cs-monomer-desorption}). This experimentally observed desorption range is broadly consistent with the thermal sublimation temperatures of $\sim$50--80~K predicted by \citet{Codella2020A&A} based on calculated H$_2$CS binding energies of 3000--4600~K on amorphous water ice. The disappearance of the monomer band is not accompanied by a dominant TPD-QMS feature suggesting that desorption is not the dominant process responsible for the loss of monomeric H$_2$CS from the ice. Instead, the temperature increase likely improves the mobility of the H$_2$CS monomer leading to their oligomerization during warm-up. In addition, desorption of the remaining H$_2$S matrix starting from 70~K may facilitate encounters between H$_2$CS monomers, thereby promoting oligomerization. In contrast, the integrated IR band at 740~cm$^{-1}$ increases in the same temperature range and persists up to ~125 K. Its disappearance coincides with the desorption feature centered near 150~K in the TPD-QMS data, supporting a contribution from the corresponding H$_2$CS-derived species to this high-temperature desorption feature. It is important to note that because the 740~cm$^{-1}$ IR band may be shared by different (H$_2$CS)$_x$ species, this correspondence does not permit an unambiguous assignment to a specific oligomer. The two weaker desorption features centered at approximately 190 and 220~K may originate from larger or more strongly bound H$_2$CS-derived aggregates.

\subsection{\texorpdfstring{Production yields of H$_2$CS and CH$_3$SH}{Production yields of H2CS and CH3SH}}

\begin{figure}
    \centering
\includegraphics[width=0.46\textwidth]{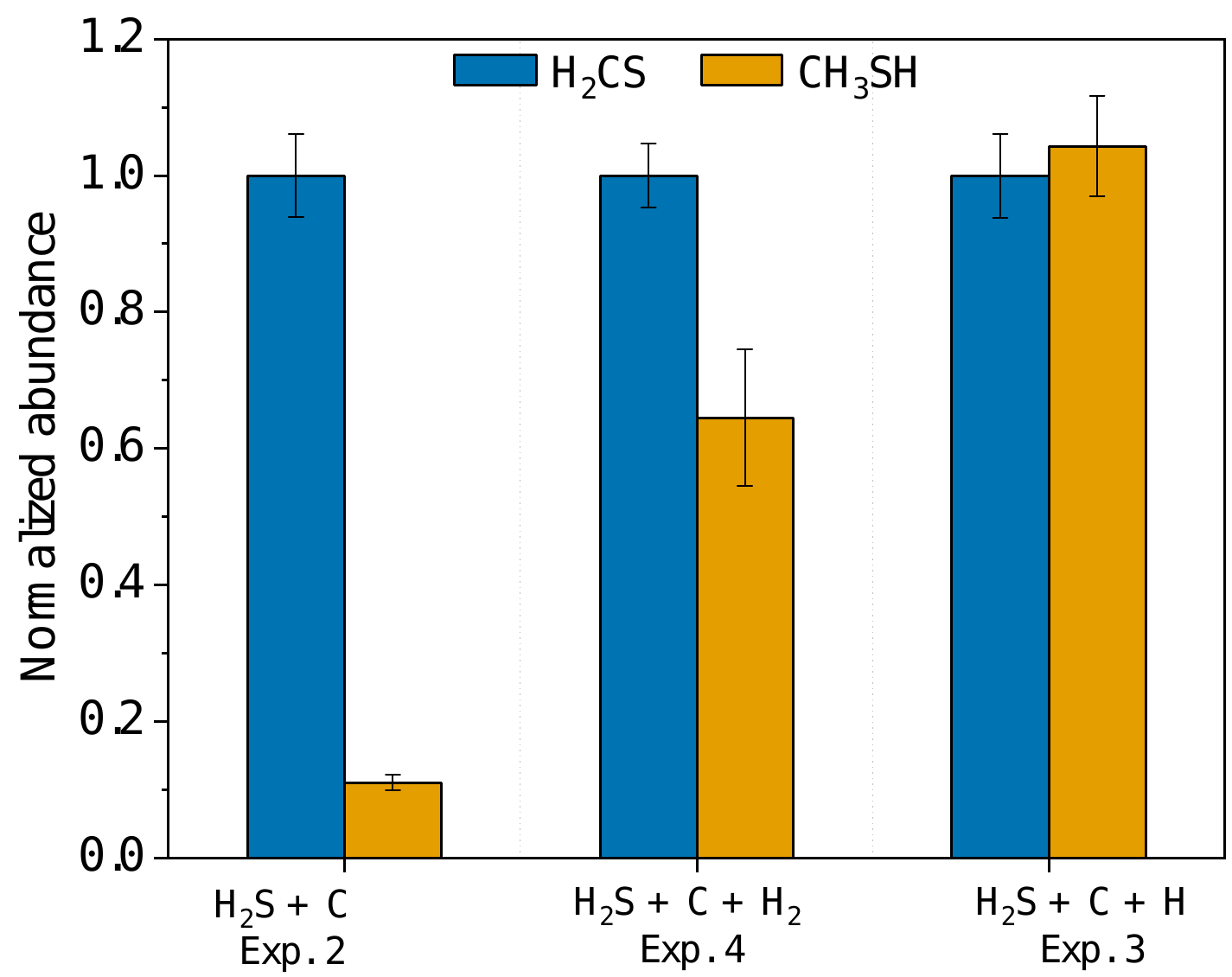}
    \caption{
Abundances of H$_2$CS and CH$_3$SH from TPD-QMS data under varying hydrogenation conditions, normalized to the H$_2$CS abundance of each respective experiment. Conditions are ordered from low to high H/H$_2$ availability. Error bars reflect uncertainties from peak integration and fitting.
}
\label{fig:abundance}
\end{figure}

Figure~\ref{fig:abundance} compares the abundance of CH$_3$SH relative to H$_2$CS for Exps.~2--4, calculated from the TPD-QMS data using Eq.~\ref{eq:CH3SH_H2CS_ratio}. For each experiment, the H$_2$CS abundance is normalized to unity, allowing the relative production of CH$_3$SH under the different hydrogenation conditions to be compared.

In the H$_2$S + C experiment (Exp.~2), CH$_3$SH is detected at only approximately 10\% of the H$_2$CS abundance. The formation of a small amount of CH$_3$SH in this experiment may result from residual H$_2$ in the chamber, which is the main component of the background gas under UHV conditions. Its detection in the TPD-QMS data, despite the absence of corresponding bands in the IR spectrum, reflects the higher sensitivity of the QMS measurements.

Upon co-deposition of H$_2$ in the H$_2$S + C + H$_2$ experiment (Exp.~4), the CH$_3$SH/H$_2$CS abundance ratio increases substantially to approximately 0.65. This shows that the increased availability of H$_2$ promotes CH$_3$SH formation, although H$_2$CS remains the dominant sulfur-bearing product under these conditions. The highest relative CH$_3$SH abundance is observed in the H$_2$S + C + H experiment (Exp.~3), where CH$_3$SH becomes as abundant as H$_2$CS. Reactive H atoms can process carbon- and sulfur-bearing intermediates toward more saturated products. At the same time, H$_2$ can also be present due to undissociated H$_2$ from the H-atom beam and in-situ association of H atoms on the surface. The high CH$_3$SH/H$_2$CS ratio may therefore contain contributions from both H-atom-driven hydrogenation and H$_2$-mediated chemistry, which cannot be disentangled from the present data.

\section{Discussion}

The obtained experimental results show the possible solid-state chemistry on dust grains involving astronomically relevant and abundant species, driving the transformation of H$_2$S ice into organosulfurs at 10 K. We demonstrate the efficient formation of H$_2$CS and its oligomers, as well as the fully saturated species CH$_3$SH, when C atoms accrete onto surfaces together with H$_2$S and H or H$_2$ at 10 K.

The formation of H$_2$CS has previously been discussed in the literature via gas-phase and grain-surface chemistry \citep[e.g.,][]{Kaiser1997JChPh, Deeyamulla2006,Lamberts2018A&A, DiGenova2026A&A}. In grain-surface networks, one proposed route involves successive H-atom addition to CS:

\begin{equation}
\mathrm{CS}
\xrightarrow{+\mathrm{H}}
\mathrm{HCS}
\xrightarrow{+\mathrm{H}}
\mathrm{H_2CS}
\end{equation}

In the theoretical network of \citet{Lamberts2018A&A}, which was largely treated without an explicit grain surface, H + CS $\rightarrow$ HCS proceeds with a small gas-phase activation barrier of 3.6~kJ~mol$^{-1}$. The subsequent H + HCS reaction is barrierless and can yield H$_2$CS as well as cis- and trans-HCSH, while H-abstraction regenerating CS also constitutes a competing channel. \citet{Lamberts2018A&A} additionally showed that the H + CS barrier changes only modestly when calculated on small water clusters. More recently, \citet{DiGenova2026A&A} investigated the hydrogenation network explicitly on water-ice clusters. The calculated H + CS barrier changes only modestly upon adsorption, and inclusion of ZPE substantially lowers the effective surface barrier. Their kinetic calculations further indicate that tunneling enables efficient CS hydrogenation at 10--20~K.

In addition to sequential hydrogenation of CS, the present experiments demonstrate that H$_2$CS can be formed from the interactions between atomic carbon and H$_2$S in the solid state. The net reaction may be represented as

\begin{equation}
\mathrm{H_2S} + \mathrm{C} \xrightarrow{} \mathrm{H_2CS}.
\label{eq:C_H2S}
\end{equation}

While the experiments confirm that the reaction occurs, they do not constrain the underlying microscopic mechanism.
Gas-phase studies suggest that C($^3$P) + H$_2$S can form an HCSH intermediate, either through initial addition of C to sulfur followed by H-atom migration \citep{Kaiser1997JChPh, Kaiser1999JChPh}, or through direct insertion of C into an S--H bond \citep{Galland2001JPCA}. In the gas phase, HCSH mainly dissociates into HCS + H, whereas stabilization by the ice matrix may allow it to rearrange into H$_2$CS \citep{Kaiser1999JChPh}. The potential role of the ice matrix in facilitating this rearrangement is illustrated by the analog H$_2$O + C$\rightarrow$ H$_2$CO system studied by \citet{Molpeceres2021JPCL}, for which carbon insertion followed by ice-assisted isomerization was found to provide an efficient route to H$_2$CO. However, the hydrogen-bond-mediated mechanism proposed for H$_2$O-rich ice cannot be assumed to operate identically in pure H$_2$S ice, and further theoretical calculations on the underlying H$_2$S + C system in the solid state are required.

When reactive H atoms are available, such as in the case of the H$_2$S + C + H experiments (Exp. 3), newly formed H$_2$CS can be further processed toward more saturated sulfur-bearing species.

\begin{equation}
\label{eq:h2cs-hydrogenation}
\mathrm{H_2CS}
\xrightarrow{+\mathrm{H}}
\mathrm{CH_2SH}
\xrightarrow{+\mathrm{H}}
\mathrm{CH_3SH}
\end{equation}

The full H-addition and H-abstraction network linking H$_2$CS to CH$_3$SH was investigated in the gas-phase by \citet{Lamberts2018A&A}. More recently, \citet{DiGenova2026A&A} extended this network to water-ice cluster models and showed that the successive H-addition reactions remain efficient at 10--20~K when tunneling is included. \citet{Lamberts2018A&A} showed that H addition to H$_2$CS preferentially forms CH$_2$SH, with an activation barrier of 3.1~kJ~mol$^{-1}$, whereas the competing H + H$_2$CS $\rightarrow$ CH$_3$S channel has a higher barrier of 7.5~kJ~mol$^{-1}$. More recently, \citet{DiGenova2026A&A} extended this network to water-ice cluster models and likewise found CH$_2$SH to be the kinetically preferred product. Their kinetic calculations further show that H-atom tunneling allows these hydrogen-addition reactions to remain efficient at 10--20~K. Once CH$_2$SH or CH$_3$S is formed, further reaction with H can produce CH$_3$SH.

In the H$_2$S + C + H$_2$ (Exp. 4) experiment, the successive hydrogenation via Eq. \ref{eq:h2cs-hydrogenation} cannot proceed due to the absence of H atoms. Alternatively, the active atomic carbon can react with partially trapped H$_2$, resulting in the formation of triplet CH$_2$ intermediates through the surface reaction:

\begin{equation}
\mathrm{C} + \mathrm{H_2} \rightarrow \mathrm{CH_2}.
\label{eq:CH2}
\end{equation}

Laboratory and computational work by \citet{Krasnokutski2016ApJ} showed that Eq.~\eqref{eq:CH2} is barrierless in helium nanodroplets, and \citet{Simoncic2020A&A} accordingly treated it as barrierless in their astrochemical models. However,
\citet{Lamberts2022ApJ} obtained activation barriers of 27--30 kJ mol$^{-1}$ for a C--H$_2$O complex and approximately 77 kJ mol$^{-1}$ for a C--(H$_2$O)$_3$ complex. The difference arises from the distinct stabilization of the reactant and transition state by the local water environment, demonstrating that the barrier is sensitive to the C-atom binding site and ice structure.
Using a larger (H$_2$O)$_{20}$ cluster, \citet{Ferrero2024ApJ} similarly found that C chemisorbs as $^3\mathrm{C{-}OH_2}$ and calculated a barrier of 70.3 kJ mol$^{-1}$ for its reaction with H$_2$. Together, these results indicate that the efficiency of Eq.~\eqref{eq:CH2} depends strongly on the adsorption state and local environment of the C atom.

The newly formed CH$_2$ radicals have been suggested to be chemically reactive and to play an essential role in ice chemistry \citep{Jin2020ApJS, Garrod2022ApJ, Simoncic2020A&A}. Hydrogen abstraction from H$_2$ is expected to be barrier-mediated (49 kJ mol$^{-1}$) and thus inefficient under the experimental conditions studied \citep{Lamberts2022ApJ}, while the H-abstraction from H$_2$S is expected to be more favorable, consistent with the efficient H-atom abstraction from H$_2$S observed on icy surfaces, which proceeds through quantum tunneling across a barrier of approximately 12.7 kJ mol$^{-1}$ \citep{Lamberts2017JPCA, Oba2018NatAs, Santos2023A&A}. CH$_2$ may therefore trigger H-abstraction from H$_2$S via

\begin{equation}
\mathrm{CH_2} + \mathrm{H_2S} \rightarrow \mathrm{CH_3} + \mathrm{SH},
\label{eq:h2s-abstraction}
\end{equation}

\noindent followed by radical recombination,
\noindent
\noindent \begin{equation}
\mathrm{CH_3} + \mathrm{SH} \rightarrow \mathrm{CH_3SH}.
\label{eq:ch3-sh-recombination}
\end{equation}

Importantly, in this mechanism CH$_3$SH formation does not require prior production and subsequent hydrogenation of H$_2$CS, but instead depends on the availability of CH$_2$ intermediates.

\subsection{Computational results}

To further assess the feasibility of this CH$_2$-mediated pathway and to identify which reaction channel is the most favorable, quantum chemical calculations were performed for the relevant CH$_3$SH formation routes.

\begin{figure*}
    \centering
    \includegraphics[width=\textwidth]{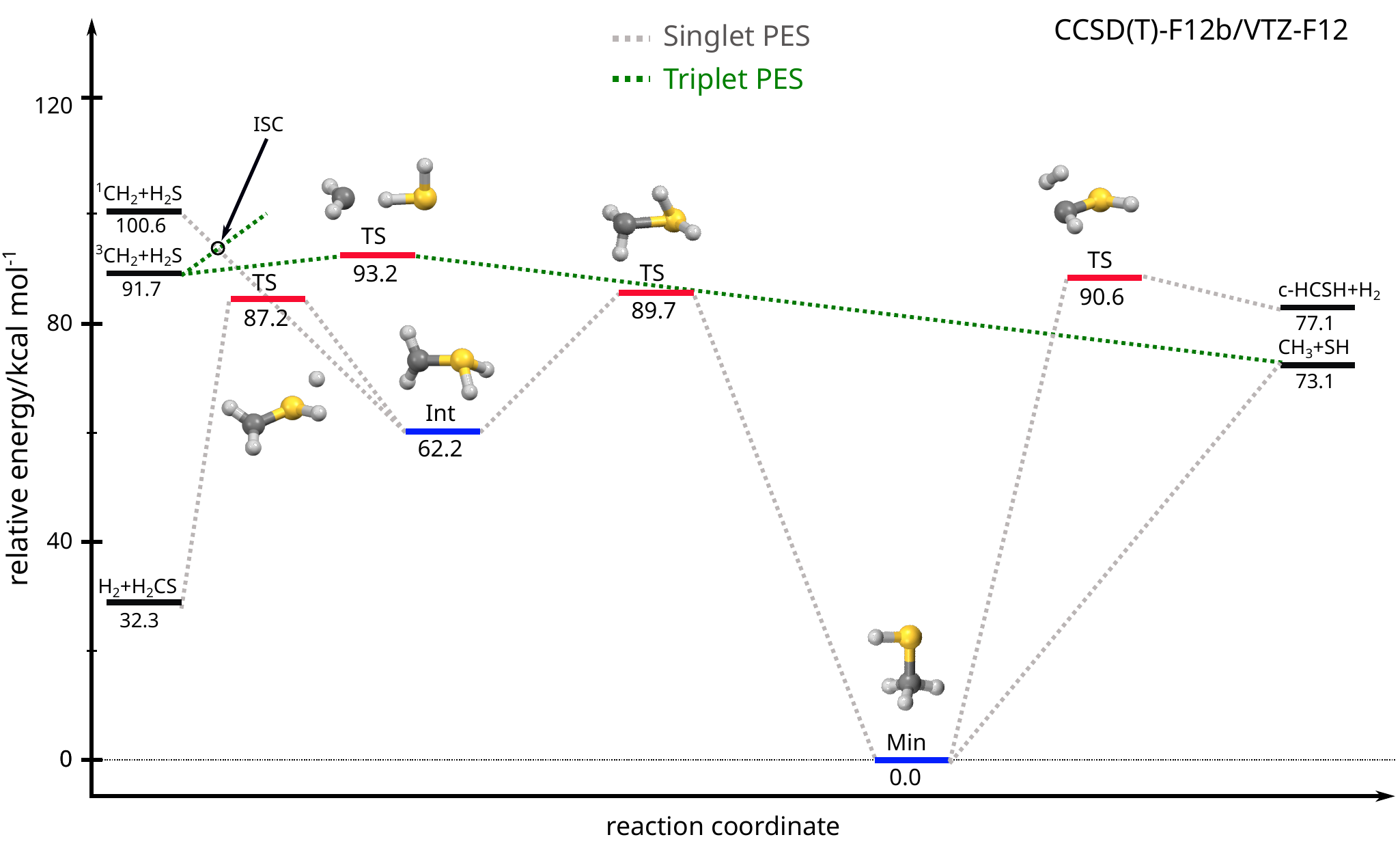}
    \caption{
    Stationary points on the ground-state singlet (gray) and triplet (green) potential energy surfaces of the CH$_3$SH system and their conversion pathways. Minima and transition states are indicated with blue and red traces, respectively, while the dissociation channels are shown in black. Optimized structures and energies were obtained at CCSD(T)-F12b/VTZ-F12 level of theory; see text for details. Energies (in kcal mol$^-1$) are given relative to the global CH$_3$SH minimum and include harmonic zero-point energy corrections. ISC stands for intersystem crossing.}
    \label{fig:ch3sh_abinitio_path}
\end{figure*}

Figure~\ref{fig:ch3sh_abinitio_path} shows the calculated stationary points for three reaction channels relevant to CH$_3$SH formation: $^1$CH$_2$ + H$_2$S, $^3$CH$_2$ + H$_2$S, and H$_2$ + H$_2$CS. All energies are reported relative to the CH$_3$SH minimum and include ZPE corrections. The H$_2$ + H$_2$CS channel is also included as an alternative pathway connecting H$_2$CS to the CH$_3$SH minimum; however, the high-lying transition states along this pathway indicate that the direct conversion of H$_2$CS by H$_2$ is unlikely to be efficient.

For the CH$_2$ + H$_2$S reaction, the singlet and triplet entrance channels are located at 100.6 and 91.7~kcal~mol$^{-1}$, respectively. The triplet pathway proceeds through a transition state at 93.2~kcal~mol$^{-1}$, corresponding to a barrier of only 1.5~kcal~mol$^{-1}$ with respect to $^3$CH$_2$ + H$_2$S, and leads to the CH$_3$ + SH radical pair at 73.1~kcal~mol$^{-1}$.

At low temperatures of the ice surface, the formation of CH$_3$SH from CH$_2$ + H$_2$S is likely dominated by the triplet abstraction pathway. The $^3$CH$_2$ + H$_2$S channel shows a very small activation barrier, allowing efficient formation of CH$_3$ and SH radicals under cold conditions. These can then recombine and be stabilized within the ice matrix toward CH$_3$SH formation. In contrast, an intersystem crossing (ISC) near the entrance channel provides possible access to the singlet PES but does not lead directly to CH$_3$SH. Instead, the singlet pathway proceeds through the CH$_2$SH$_2$ intermediate, which must overcome a transition state at 89.7~kcal~mol$^{-1}$, corresponding to a barrier of 27.5~kcal~mol$^{-1}$ relative to the intermediate. The competition between ISC and the rapid barrier crossing along the triplet pathway, together with the additional barrier separating CH$_2$SH$_2$ from CH$_3$SH, makes the singlet route unlikely to compete with triplet H-abstraction followed by radical recombination.

In the context of the present experiments, the above gas-phase quantum chemistry calculations support a mechanism in which CH$_2$, formed through the reaction of C atoms with H$_2$, reacts with H$_2$S mainly via triplet H-abstraction, yielding CH$_3$ and SH radicals that subsequently recombine and are stabilized as CH$_3$SH within the ice matrix.

\subsection{\texorpdfstring{Proposed pathways to H$_2$CS and CH$_3$SH}{Proposed pathways to H2CS and CH3SH}}

\begin{figure}
    \centering \includegraphics[width=\columnwidth]{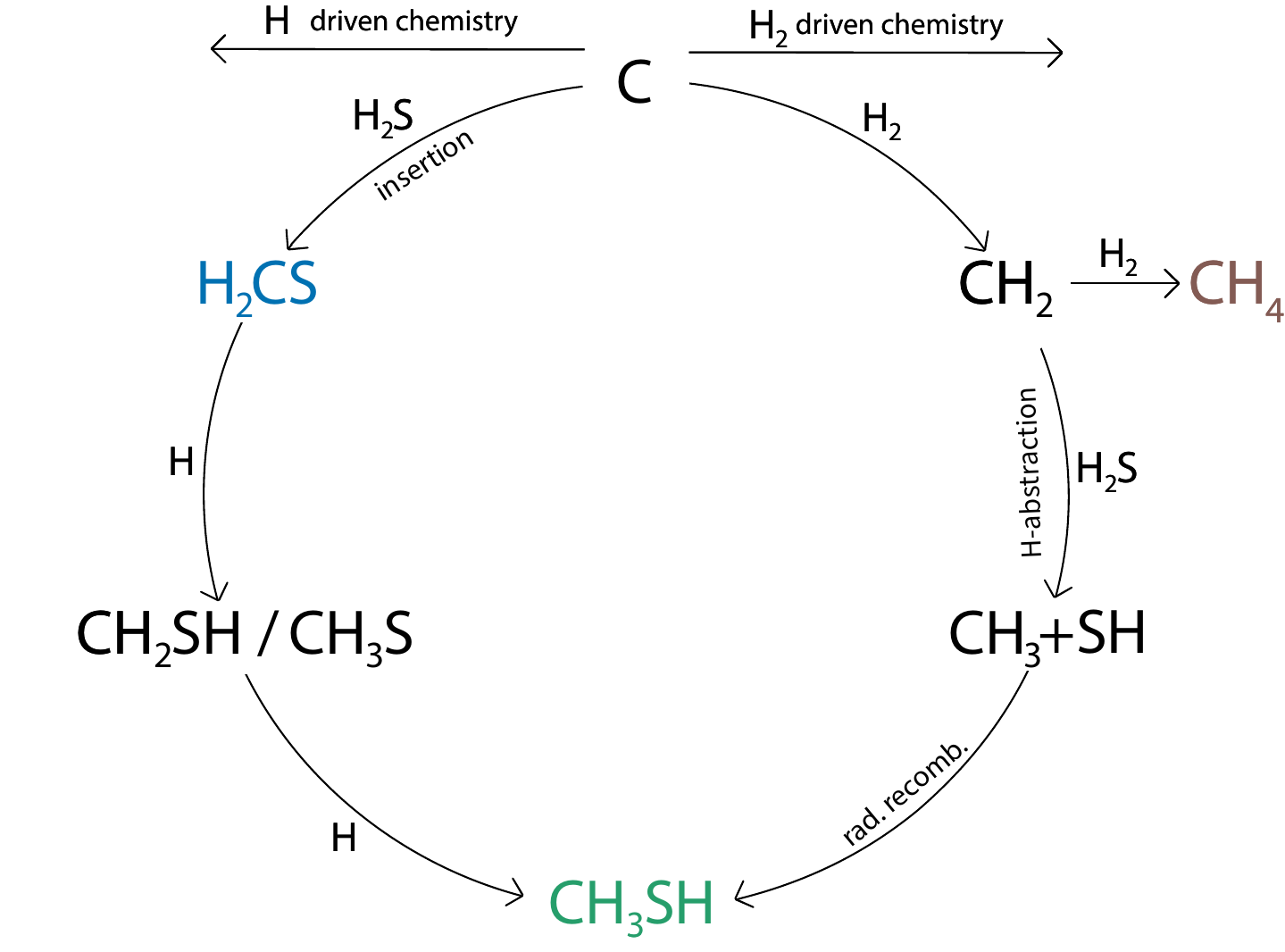}
    \caption{
Proposed reaction network for the formation of H$_2$CS and CH$_3$SH during co-deposition experiments involving C atoms, H atoms, and H$_2$.}
    \label{fig:formation_network}
\end{figure}

Figure~\ref{fig:formation_network} summarizes the reaction network inferred from the present experiments and highlights that sulfur chemistry in H$_2$S-rich ices at 10~K can proceed via two complementary but mechanistically distinct regimes. This depends on whether reactive H atoms or only H$_2$ molecules are available along with accreting C atoms.

In the H-driven regime, atomic C reacts directly with H$_2$S to form H$_2$CS, most plausibly through carbon insertion into the S--H bond, i.e., Eq.~\eqref{eq:C_H2S}. Once formed, H$_2$CS becomes an intermediate that can undergo successive H-atom addition reactions to yield CH$_2$SH/CH$_3$S radicals and ultimately CH$_3$SH. In this branch, H$_2$CS is therefore both a primary product of the C + H$_2$S chemistry and the precursor to more saturated sulfur-bearing species.

In the H$_2$-driven regime, CH$_3$SH formation cannot be explained by direct hydrogenation of H$_2$CS, because reactive H atoms are absent. Instead, the key additional step is the formation of CH$_2$ intermediates through the reaction of C with H$_2$ (Eq.~\ref{eq:CH2}). The next step involves H-abstraction from H$_2$S to form CH$_3$ and SH radicals (Eq.~\ref{eq:h2s-abstraction}), followed by radical recombination to produce CH$_3$SH (Eq.~\ref{eq:ch3-sh-recombination}). In this branch, CH$_3$SH formation proceeds independently of H$_2$CS and instead depends on the availability of CH$_2$ produced by C + H$_2$ chemistry. The coexistence of H$_2$CS, CH$_3$SH and CH$_4$ therefore reflects competition between insertion, hydrogenation, abstraction and radical--radical recombination processes on the ice surface.

The gas-phase calculations provide a reference for assessing the relative energies and activation barriers of the proposed reaction pathways. Previous calculations for H + CS $\rightarrow$ HCS on two water clusters \citep{Lamberts2018A&A} and H + H$_2$S $\rightarrow$ H$_2$ + HS in the presence of H$_2$O or H$_2$S clusters \citep{Lamberts2017JPCA} found relatively small changes in activation energies compared with the corresponding gas-phase reactions, supporting the use of this approach as a first approximation. In the present experiments, the ice environment may additionally influence reactant orientations and facilitate stabilization of the reaction products through energy dissipation. Although the magnitude of these effects has not been evaluated explicitly for CH$_2$ + H$_2$S, the calculations support the proposed reaction pathway leading to the observed CH$_3$SH formation.

\section{Astrophysical implications}

This laboratory work provides new experimental constraints on grain-surface reaction routes leading to the sulfur-bearing organic molecules H$_2$CS and CH$_3$SH. Although these species are observed in the gas phase, their grain-surface formation routes have only been partially explored theoretically and require systematic experimental studies. Therefore, the present experiments target chemical processes that may operate in environments where atomic carbon, molecular hydrogen, and sulfur-bearing species such as H$_2$S can coexist on grain surfaces during the earliest stages of star formation.

Neutral C atoms are abundantly present in translucent clouds, typically associated with visual extinctions of $A_V\sim1-5$ mag, with observed column-averaged abundances reaching C/H$_2$ $\sim2.2\times10^{-5}$ at intermediate extinctions \citep{vanDishoeck1989ApJ, Snow2006ARA&A}.
Additionally, observational and theoretical studies have suggested that atomic carbon may remain spatially extended, possibly even to denser regions \citep{Langer1976ApJ, Keene1985ApJ, Zmuidzinas1988ApJ, Papadopoulos2004MNRAS, Burton2015ApJ, Bisbas2019MNRAS}. In addition, atomic carbon can become available again in cold prestellar cores, where residual gas-phase CO can be destroyed by He$^+$, producing C$^+$ that is further neutralized to enrich neutral atomic carbon \citep{Ruffle1999MNRAS,Bialy2015MNRAS}. Therefore, C-atom accretion and subsequent solid-state chemistry are highly expected during the onset of ice growth and at later prestellar stages.

Ultraviolet absorption studies indicate that sulfur abundance is only weakly depleted along diffuse and translucent sight lines \citep{Snow1996ApJ, Jenkins2009ApJ}, indicating that little to no sulfur has yet been incorporated into icy mantles at this early,
low-density stage. Observations of singly and doubly deuterated H$_2$S toward Class~0 sources indicate that H$_2$S formation on grains predates the onset of catastrophic CO freeze-out \citep{Ceccarelli2014prpl}, which occurs only later, at the higher densities reached in prestellar cores. Taken together, these constraints suggest that H$_2$S formation does not begin until the cloud has evolved beyond the translucent stage into the denser molecular gas that precedes prestellar core formation, at which point atomic carbon is still available for accretion onto H$_2$S-bearing ices before it is locked into CO.

In dense molecular clouds, H atoms accrete onto grain surfaces and diffuse to react with available species, including H$_2$S. When C atoms instead participate in reactions with H$_2$S, they form H$_2$CS, which can subsequently be hydrogenated by H atoms to yield CH$_3$SH. However, H$_2$ is several orders of magnitude more abundant than atomic H in these environments \citep{Snow2006ARA&A}, and despite its comparatively low sticking coefficient, H$_2$ is expected to also participate in solid-state reactions on grain surfaces. In particular, C + H$_2$ $\rightarrow$ CH$_2$ has been proposed to occur on grain surfaces, and the resulting CH$_2$ can react with H$_2$S to form CH$_3$SH via a two-step process involving H-abstraction followed by radical recombination. These two routes -- H-atom-driven hydrogenation of H$_2$CS and H$_2$-mediated formation via CH$_2$ -- can therefore operate simultaneously, with the relative importance of each mechanism depending on the local physical conditions. The present experiments show that this distinction matters, since CH$_3$SH formation can occur through both pathways.

The distinction between H-atom-driven and H$_2$-mediated chemistry may also have implications for isotopic fractionation. In prestellar cores, gas-phase deuterium fractionation enhances the H$_2$D$^+$/H$_3^+$ ratio well above the elemental D/H ratio \citep{Watson1974ApJ}, and this fractionation is subsequently transferred to reactive atomic and molecular species participating in grain-surface chemistry. By contrast, the bulk H$_2$/HD reservoir remains close to the elemental D/H ratio, so pathways that use H$_2$ directly as a reactant are, in principle, less likely to introduce strong deuterium enrichment than pathways relying on sequential H- and D-atom addition. Testing this expectation, however, requires a direct comparison of the reaction efficiencies of H versus D and of H$_2$ versus D$_2$ in these systems, which we identify as a valuable direction for future experimental work.

Recent astronomical observations also argue against a simple direct analogy between the hydrogenation of H$_2$CS to CH$_3$SH and that of H$_2$CO to CH$_3$OH \citep{Bunn2025ApJ}. In the prestellar core L1544, Bunn et al.\ report a lower limit of H$_2$CS/CH$_3$SH $>2760$, which is already about an order of magnitude higher than the corresponding oxygen-bearing ratio H$_2$CO/CH$_3$OH $= 282.3 \pm 177.2$. This large difference may reflect a combination of factors, such as efficient H$_2$CS formation through the reaction of C atoms with H$_2$S, limited availability of H atoms to drive its further hydrogenation, or efficient destruction of CH$_3$SH once formed. The present experiments identify both H-atom-driven and H$_2$-mediated routes to CH$_3$SH in the laboratory, and further observational constraints on the H$_2$CS/CH$_3$SH ratio across a range of sources will help clarify their relative contributions under interstellar conditions.

Together, these results indicate that CH$_3$SH formation in interstellar ices need not rely solely on sequential H-atom addition, but can also proceed through H$_2$-mediated pathways initiated by accreting C atoms.

The efficiency of the reaction routes identified here under interstellar ice conditions will depend on several factors that are not captured directly by the present experiments, including competition with other surface reactions, the binding and diffusion energies of the reactants, and the influence of local binding environments and molecular orientation on reaction probabilities \citep[e.g.,][]{EnriqueRomero2021A&A,Tsuge2023NatAs,Ferrero2024ApJ, Martinez-Bachs2026ApJ}. Astrochemical modeling incorporating these factors will be needed to simulate the full network of surface chemistry induced by accreting C atoms in realistic astronomical environments and to assess the relative importance of the pathways identified in this work.

\section{Conclusions}

We present a systematic laboratory study of the formation of sulfur-bearing organic molecules induced by reactions of C atoms with H$_2$S ice, in the absence and presence of H$_2$ or H atoms on grain surfaces at 10~K. By combining RAIRS and TPD measurements, we securely identify the formation of H$_2$CS and CH$_3$SH and propose plausible reaction pathways for their production under dense molecular cloud conditions. Our main conclusions are summarized as follows:

\begin{enumerate}[label=({\roman*})]

\item The interaction of atomic carbon with H$_2$S at 10~K efficiently forms H$_2$CS. The IR spectra reveal H$_2$CS in the forms of monomer and the H$_2$CS-derived oligomers, (H$_2$CS)$_x$. The reaction may proceed through initial insertion of atomic carbon into an S--H bond, followed by rearrangement to H$_2$CS.

\item CH$_3$SH forms via hydrogenation of H$_2$CS as demonstrated in the H$_2$S + C + H experiments. The IR spectra at 10~K show characteristic CH$_3$SH bands in the investigated spectral range, while the TPD-QMS data confirm a CH$_3$SH desorption peak at $\sim$ 110~K by examining the corresponding fragmentation pattern with literature data.

\item An additional formation route to CH$_3$SH is studied in the H$_2$S + C + H$_2$ experiment via the CH$_2$-induced H-abstraction reaction of H$_2$S, followed by CH$_3$ + SH radical recombination. Theoretical calculations suggest that the CH$_2$ + H$_2$S $\rightarrow$ CH$_3$ + SH reaction is possible, with a relatively low barrier of 1.5 kcal.mol$^{-1}$ while the radical--molecule association pathway (i.e., CH$_2$ + H$_2$S $\rightarrow$ CH$_3$SH) is not favored.

\end{enumerate}

The experimental results suggest efficient reactions of H$_2$S with C atoms and CH$_2$ radicals on grain surfaces at 10 K, forming the organosulfurs H$_2$CS and CH$_3$SH. This surface chemistry enriches the interstellar sulfur reaction network and hints at an underexplored route to sulfur-bearing molecular complexity in ices. Given the large abundance of H$_2$ relative to H atoms, newly formed CH$_2$ radicals produced via the C + H$_2$ reaction may open additional reaction channels that enhance the formation of complex molecules in interstellar ices. Future astrochemical modeling that accounts for the reaction network identified here will be essential to determine when carbon-driven and H$_2$-mediated chemistry can efficiently contribute to sulfur-bearing organic formation in interstellar ices.

\begin{acknowledgements}

F.K. acknowledges support from the European Union’s Horizon Europe research and innovation program under the Marie Skłodowska-Curie Actions Postdoctoral Fellowship grant agreement No. \#~$101153804$ (ORCHID -- Organic Reactions of Carbon with H$_2$ in Interstellar Dust). F.K. thanks Paola Caselli, Silvia Spezzano, and Olli Sipilä for insightful discussions on the astronomical implications of this work. This work has been funded by NOVA and NWO-DANIII. J.C.S. was supported by the Heising-Simons Foundation through a 51 Pegasi b Fellowship. S.I. acknowledges support from the Danish National Research Foundation through the Centre of Excellence ‘InterCat’ (grant agreement No. DNRF150).

\end{acknowledgements}

\bibliographystyle{aa}

\bibliography{Refs_ORCHID}

@ARTICLE{Watson1974ApJ,
       author = {{Watson}, William D.},
        title = "{Ion-Molecule Reactions, Molecule Formation, and Hydrogen-Isotope Exchange in Dense Interstellar Clouds}",
      journal = {\apj},
         year = 1974,
        month = feb,
       volume = {188},
        pages = {35-42},
          doi = {10.1086/152681},
       adsurl = {https://ui.adsabs.harvard.edu/abs/1974ApJ...188...35W}
}

@ARTICLE{Jacox1975JMoSp,
       author = {{Jacox}, Marilyn E. and {Milligan}, Dolphus E.},
        title = "{Matrix isolation study of the infrared spectrum of thioformaldehyde}",
      journal = {J. Mol. Spectrosc.},
         year = 1975,
        month = oct,
       volume = {58},
       number = {1},
        pages = {142-157},
          doi = {10.1016/0022-2852(75)90162-9},
       adsurl = {https://ui.adsabs.harvard.edu/abs/1975JMoSp..58..142J}
}

@ARTICLE{Langer1976ApJ,
       author = {{Langer}, W.},
        title = "{The carbon monoxide abundance in interstellar clouds.}",
      journal = {\apj},
         year = 1976,
        month = jun,
       volume = {206},
        pages = {699-712},
          doi = {10.1086/154430},
       adsurl = {https://ui.adsabs.harvard.edu/abs/1976ApJ...206..699L}
}

@ARTICLE{Keene1985ApJ,
       author = {{Keene}, J. and {Blake}, G.~A. and {Phillips}, T.~G. and {Huggins}, P.~J. and {Beichman}, C.~A.},
        title = "{The abundance of atomic carbon near the ionization fronts in M17 and S140.}",
      journal = {\apj},
         year = 1985,
        month = dec,
       volume = {299},
        pages = {967-980},
          doi = {10.1086/163763},
       adsurl = {https://ui.adsabs.harvard.edu/abs/1985ApJ...299..967K}
}

@ARTICLE{Zmuidzinas1988ApJ,
       author = {{Zmuidzinas}, J. and {Betz}, A.~L. and {Boreiko}, R.~T. and {Goldhaber}, D.~M.},
        title = "{Neutral Atomic Carbon in Dense Molecular Clouds}",
      journal = {\apj},
         year = 1988,
        month = dec,
       volume = {335},
        pages = {774},
          doi = {10.1086/166966},
       adsurl = {https://ui.adsabs.harvard.edu/abs/1988ApJ...335..774Z}
}

@ARTICLE{vanDishoeck1989ApJ,
       author = {{van Dishoeck}, Ewine F. and {Black}, John H.},
        title = "{Interstellar C 2, CH, and CN in Translucent Molecular Clouds}",
      journal = {\apj},
         year = 1989,
        month = may,
       volume = {340},
        pages = {273},
          doi = {10.1086/167391},
       adsurl = {https://ui.adsabs.harvard.edu/abs/1989ApJ...340..273V}
}

@article{DUN89:1007,
author = "Dunning, Thom H.",
title = "{Gaussian Basis Sets for Use in Correlated Molecular Calculations. I. The Atoms Boron through Neon and Hydrogen}",
journal = {\jcp},
year = "1989",
volume = "90",
number = "2", 
pages = "1007-1023",
doi = "10.1063/1.456153" 
}

@ARTICLE{Smith1991MNRAS,
       author = {{Smith}, Robert G.},
        title = "{A search for solid H$_{2}$S in dense clouds.}",
      journal = {\mnras},
         year = 1991,
        month = mar,
       volume = {249},
        pages = {172},
          doi = {10.1093/mnras/249.1.172},
       adsurl = {https://ui.adsabs.harvard.edu/abs/1991MNRAS.249..172S}
}

@article{KEN92:6796,
author = "Kendall, Rick A. and Dunning, Thom H. and Harrison, Robert J.",
title = "{Electron Affinities of the First-Row Atoms Revisited. Systematic Basis Sets and Wave Functions}",
journal = {\jcp},
year = "1992",
volume = "96",
number = "9", 
pages = "6796-6806",
doi = "10.1063/1.462569" 
}

@ARTICLE{Palumbo1995ApJ,
       author = {{Palumbo}, M.~E. and {Tielens}, A.~G.~G.~M. and {Tokunaga}, A.~T.},
        title = "{Solid Carbonyl Sulphide (OCS) in W33A}",
      journal = {\apj},
         year = 1995,
        month = aug,
       volume = {449},
        pages = {674},
          doi = {10.1086/176088},
       adsurl = {https://ui.adsabs.harvard.edu/abs/1995ApJ...449..674P}
}

@ARTICLE{Snow1996ApJ,
       author = {{Snow}, Theodore P. and {Black}, John H. and {van Dishoeck}, Ewine F. and {Burks}, Geoffrey and {Crutcher}, Richard M. and {Lutz}, Barry L. and {Hanson}, Margaret M. and {Shuping}, Ralph Y.},
        title = "{The Translucent Molecular Clouds toward HD 154368. I. Extinction, Abundances, and Depletions}",
      journal = {\apj},
         year = 1996,
        month = jul,
       volume = {465},
        pages = {245},
          doi = {10.1086/177416},
       adsurl = {https://ui.adsabs.harvard.edu/abs/1996ApJ...465..245S}
}

@ARTICLE{Kaiser1997JChPh,
       author = {{Kaiser}, R.~I. and {Sun}, W. and {Suits}, A.~G.},
        title = "{Crossed beam reaction of atomic carbon C($^{3}$P$_{j}$) with hydrogen sulfide, H$_{2}$S(X$^{1}$A$_{1}$): Observation of the thioformyl radical, HCS(X$^{2}$A')}",
      journal = {\jcp},
         year = 1997,
        month = mar,
       volume = {106},
       number = {12},
        pages = {5288-5291},
          doi = {10.1063/1.473561},
       adsurl = {https://ui.adsabs.harvard.edu/abs/1997JChPh.106.5288K}
}

@ARTICLE{Kaiser1999JChPh,
       author = {{Kaiser}, R.~I. and {Ochsenfeld}, C. and {Head-Gordon}, M. and {Lee}, Y.~T.},
        title = "{Crossed-beam reaction of carbon atoms with sulfur containing molecules. I. Chemical dynamics of thioformyl (HCS X$^{2}$A$^{'}$) formation from reaction of C($^{3}$P$_{j}$) with hydrogen sulfide, H2S(X$^{1}$A$_{1}$)}",
      journal = {\jcp},
         year = 1999,
        month = feb,
       volume = {110},
       number = {5},
        pages = {2391-2403},
          doi = {10.1063/1.477944},
       adsurl = {https://ui.adsabs.harvard.edu/abs/1999JChPh.110.2391K}
}

@ARTICLE{Ruffle1999MNRAS,
       author = {{Ruffle}, D.~P. and {Hartquist}, T.~W. and {Caselli}, P. and {Williams}, D.~A.},
        title = "{The sulphur depletion problem}",
      journal = {\mnras},
         year = 1999,
        month = jul,
       volume = {306},
       number = {3},
        pages = {691-695},
          doi = {10.1046/j.1365-8711.1999.02562.x},
       adsurl = {https://ui.adsabs.harvard.edu/abs/1999MNRAS.306..691R}
}

@ARTICLE{Caselli1999ApJ,
       author = {{Caselli}, P. and {Walmsley}, C.~M. and {Tafalla}, M. and {Dore}, L. and {Myers}, P.~C.},
        title = "{CO Depletion in the Starless Cloud Core L1544}",
      journal = {ApJL},
         year = 1999,
        month = oct,
       volume = {523},
       number = {2},
        pages = {L165-L169},
          doi = {10.1086/312280},
       adsurl = {https://ui.adsabs.harvard.edu/abs/1999ApJ...523L.165C}
}

@article{Galland2001JPCA,
  author  = {Galland, Nicolas and Caralp, Fran{\c{c}}oise and
             Rayez, Marie-Th{\'e}r{\`e}se and Hannachi, Yacine and
             Loison, Jean-Christophe and Dorthe, G{\'e}rard and
             Bergeat, Astrid},
  title   = {Reaction of Carbon Atoms, {C} (2p$^2$, $^3${P}), with
             Hydrogen Sulfide, {H$_2$S} ({X} $^1${A}$_1$):
             Overall Rate Constant and Product Channels},
  journal = {J. Phys. Chem. A},
  year    = {2001},
  volume  = {105},
  number  = {43},
  pages   = {9893--9900},
  doi     = {10.1021/jp011713m}
}

@ARTICLE{Watanabe2002ApJ,
       author = {{Watanabe}, Naoki and {Kouchi}, Akira},
        title = "{Efficient Formation of Formaldehyde and Methanol by the Addition of Hydrogen Atoms to CO in H$_{2}$O-CO Ice at 10 K}",
      journal = {ApJL},
         year = 2002,
        month = jun,
       volume = {571},
       number = {2},
        pages = {L173-L176},
          doi = {10.1086/341412},
       adsurl = {https://ui.adsabs.harvard.edu/abs/2002ApJ...571L.173W}
}

@article{Hudson2003_PCCP_alcohols,
  author  = {Hudson, James E. and Hamilton, Michelle L. and Vallance, Claire and Harland, Peter W.},
  title   = {Absolute electron impact ionization cross-sections for the C1 to C4 alcohols},
  journal = {Phys. Chem. Chem. Phys.},
  year    = {2003},
  volume  = {5},
  pages   = {3162--3168},
  doi     = {10.1039/B304456D}
}

@ARTICLE{Papadopoulos2004MNRAS,
       author = {{Papadopoulos}, P.~P. and {Thi}, W.-F. and {Viti}, S.},
        title = "{CI lines as tracers of molecular gas, and their prospects at high redshifts}",
      journal = {\mnras},
         year = 2004,
        month = jun,
       volume = {351},
       number = {1},
        pages = {147-160},
          doi = {10.1111/j.1365-2966.2004.07762.x},
archivePrefix = {arXiv},
       eprint = {astro-ph/0403092},
 primaryClass = {astro-ph},
       adsurl = {https://ui.adsabs.harvard.edu/abs/2004MNRAS.351..147P}
}

@ARTICLE{Esteban2004MNRAS,
       author = {{Esteban}, C. and {Peimbert}, M. and {Garc{\'\i}a-Rojas}, J. and {Ruiz}, M.~T. and {Peimbert}, A. and {Rodr{\'\i}guez}, M.},
        title = "{A reappraisal of the chemical composition of the Orion nebula based on Very Large Telescope echelle spectrophotometry}",
      journal = {\mnras},
         year = 2004,
        month = nov,
       volume = {355},
       number = {1},
        pages = {229-247},
          doi = {10.1111/j.1365-2966.2004.08313.x},
archivePrefix = {arXiv},
       eprint = {astro-ph/0408249},
 primaryClass = {astro-ph},
       adsurl = {https://ui.adsabs.harvard.edu/abs/2004MNRAS.355..229E}
}

@ARTICLE{Snow2006ARA&A,
       author = {{Snow}, Theodore P. and {McCall}, Benjamin J.},
        title = "{Diffuse Atomic and Molecular Clouds}",
      journal = {\araa},
         year = 2006,
        month = sep,
       volume = {44},
       number = {1},
        pages = {367-414},
          doi = {10.1146/annurev.astro.43.072103.150624},
       adsurl = {https://ui.adsabs.harvard.edu/abs/2006ARA&A..44..367S}
}

@article{Deeyamulla2006,
  author  = {Deeyamulla, M. P. and Husain, D.},
  title   = {Kinetic studies of atomic carbon, C[2p2(3PJ)], with small sulfur-containing molecules by time-resolved atomic resonance absorption spectroscopy in the vacuum ultra-violet},
  journal = {J. Photochem. Photobiol. A: Chem.},
  year    = {2006},
  volume  = {184},
  pages   = {347--353},
  doi     = {10.1016/j.jphotochem.2006.04.038}
}

@ARTICLE{Suzuki2007VibSp,
       author = {{Suzuki}, Eiichi and {Yamazaki}, Masahiro and {Shimizu}, Kenji},
        title = "{Infrared spectra of monomeric thioformaldehyde in Ar, N <SUB loc=``post''>2</SUB> and Xe matrices}",
      journal = {Vib. Spectrosc.},
         year = 2007,
        month = jan,
       volume = {43},
       number = {2},
        pages = {269-273},
          doi = {10.1016/j.vibspec.2006.02.007},
       adsurl = {https://ui.adsabs.harvard.edu/abs/2007VibSp..43..269S}
}

@ARTICLE{Bergin2007ARA&A,
       author = {{Bergin}, Edwin A. and {Tafalla}, Mario},
        title = "{Cold Dark Clouds: The Initial Conditions for Star Formation}",
      journal = {\araa},
         year = 2007,
        month = sep,
       volume = {45},
       number = {1},
        pages = {339-396},
          doi = {10.1146/annurev.astro.45.071206.100404},
archivePrefix = {arXiv},
       eprint = {0705.3765},
 primaryClass = {astro-ph},
       adsurl = {https://ui.adsabs.harvard.edu/abs/2007ARA&A..45..339B}
}

@ARTICLE{Ioppolo2008ApJ,
       author = {{Ioppolo}, S. and {Cuppen}, H.~M. and {Romanzin}, C. and {van Dishoeck}, E.~F. and {Linnartz}, H.},
        title = "{Laboratory Evidence for Efficient Water Formation in Interstellar Ices}",
      journal = {\apj},
         year = 2008,
        month = oct,
       volume = {686},
       number = {2},
        pages = {1474-1479},
          doi = {10.1086/591506},
archivePrefix = {arXiv},
       eprint = {0807.0129},
 primaryClass = {astro-ph},
       adsurl = {https://ui.adsabs.harvard.edu/abs/2008ApJ...686.1474I}
}

@article{Bull2008_IJMS_polarizability,
  author  = {Bull, James N. and Harland, Peter W.},
  title   = {Absolute electron impact ionization cross-sections and polarisability volumes for C2 to C4 aldehydes, C4 and C6 symmetric ethers and C3 to C6 ketones},
  journal = {Int. J. Mass Spectrom.},
  year    = {2008},
  volume  = {273},
  pages   = {53--57},
  doi     = {10.1016/j.ijms.2008.03.003}
}

@ARTICLE{Jenkins2009ApJ,
       author = {{Jenkins}, Edward B.},
        title = "{A Unified Representation of Gas-Phase Element Depletions in the Interstellar Medium}",
      journal = {\apj},
         year = 2009,
        month = aug,
       volume = {700},
       number = {2},
        pages = {1299-1348},
          doi = {10.1088/0004-637X/700/2/1299},
archivePrefix = {arXiv},
       eprint = {0905.3173},
 primaryClass = {astro-ph.GA},
       adsurl = {https://ui.adsabs.harvard.edu/abs/2009ApJ...700.1299J}
}

@ARTICLE{Asplund2009ARA&A,
       author = {{Asplund}, Martin and {Grevesse}, Nicolas and {Sauval}, A. Jacques and {Scott}, Pat},
        title = "{The Chemical Composition of the Sun}",
      journal = {\araa},
         year = 2009,
        month = sep,
       volume = {47},
       number = {1},
        pages = {481-522},
          doi = {10.1146/annurev.astro.46.060407.145222},
archivePrefix = {arXiv},
       eprint = {0909.0948},
 primaryClass = {astro-ph.SR},
       adsurl = {https://ui.adsabs.harvard.edu/abs/2009ARA&A..47..481A}
}

@ARTICLE{Herbst2009ARA&A,
       author = {{Herbst}, Eric and {van Dishoeck}, Ewine F.},
        title = "{Complex Organic Interstellar Molecules}",
      journal = {\araa},
         year = 2009,
        month = sep,
       volume = {47},
       number = {1},
        pages = {427-480},
          doi = {10.1146/annurev-astro-082708-101654},
       adsurl = {https://ui.adsabs.harvard.edu/abs/2009ARA&A..47..427H}
}

@ARTICLE{Fuchs2009A&A,
       author = {{Fuchs}, G.~W. and {Cuppen}, H.~M. and {Ioppolo}, S. and {Romanzin}, C. and {Bisschop}, S.~E. and {Andersson}, S. and {van Dishoeck}, E.~F. and {Linnartz}, H.},
        title = "{Hydrogenation reactions in interstellar CO ice analogues. A combined experimental/theoretical approach}",
      journal = {\aap},
         year = 2009,
        month = oct,
       volume = {505},
       number = {2},
        pages = {629-639},
          doi = {10.1051/0004-6361/200810784},
       adsurl = {https://ui.adsabs.harvard.edu/abs/2009A&A...505..629F}
}

@article{KNI09:054104,
    author = {Knizia, Gerald and Adler, Thomas B. and Werner, Hans-Joachim},
    title = {Simplified CCSD(T)-F12 methods: Theory and benchmarks},
    journal = {\jcp},
    volume = {130},
    number = {5},
    pages = {054104},
    year = {2009},
    doi = {10.1063/1.3054300},
}

@ARTICLE{Dulieu2010A&A,
       author = {{Dulieu}, F. and {Amiaud}, L. and {Congiu}, E. and {Fillion}, J.-H. and {Matar}, E. and {Momeni}, A. and {Pirronello}, V. and {Lemaire}, J.~L.},
        title = "{Experimental evidence for water formation on interstellar dust grains by hydrogen and oxygen atoms}",
      journal = {\aap},
         year = 2010,
        month = mar,
       volume = {512},
          eid = {A30},
        pages = {A30},
          doi = {10.1051/0004-6361/200912079},
archivePrefix = {arXiv},
       eprint = {0903.3120},
 primaryClass = {astro-ph.IM},
       adsurl = {https://ui.adsabs.harvard.edu/abs/2010A&A...512A..30D}
}

@ARTICLE{JimezesEscobar2011A&A,
       author = {{Jim{\'e}nez-Escobar}, A. and {Mu{\~n}oz Caro}, G.~M.},
        title = "{Sulfur depletion in dense clouds and circumstellar regions. I. H$_{2}$S ice abundance and UV-photochemical reactions in the H$_{2}$O-matrix}",
      journal = {\aap},
         year = 2011,
        month = dec,
       volume = {536},
          eid = {A91},
        pages = {A91},
          doi = {10.1051/0004-6361/201014821},
archivePrefix = {arXiv},
       eprint = {1112.3240},
 primaryClass = {astro-ph.EP},
       adsurl = {https://ui.adsabs.harvard.edu/abs/2011A&A...536A..91J}
}

@article{SZA012:108,
author = {Szalay, P\'{e}ter G. and M\"{u}ller, Thomas and Gidofalvi, Gergely and Lischka, Hans and Shepard, Ron},
title = "{Multiconfiguration Self-Consistent Field and Multireference Configuration Interaction Methods and Applications}",
journal = {Chem. Rev.},
volume = {112},
number = {1},
pages = {108-181},
year = {2012},
doi = {10.1021/cr200137a},
}

@ARTICLE{Ioppolo2013RScI,
       author = {{Ioppolo}, S. and {Fedoseev}, G. and {Lamberts}, T. and {Romanzin}, C. and {Linnartz}, H.},
        title = "{SURFRESIDE$^{2}$: An ultrahigh vacuum system for the investigation of surface reaction routes of interstellar interest}",
      journal = {Rev. Sci. Instrum.},
         year = 2013,
        month = jul,
       volume = {84},
       number = {7},
          eid = {073112-073112-13},
        pages = {073112-073112-13},
          doi = {10.1063/1.4816135},
       adsurl = {https://ui.adsabs.harvard.edu/abs/2013RScI...84g3112I}
}

@article{FEL013:084110,
    author = {Feller, David and Peterson, Kirk A.},
    title = {An expanded calibration study of the explicitly correlated CCSD(T)-F12b method using large basis set standard CCSD(T) atomization energies},
    journal = {\jcp},
    volume = {139},
    number = {8},
    pages = {084110},
    year = {2013},
    doi = {10.1063/1.4819125},
}

@INPROCEEDINGS{Ceccarelli2014prpl,
       author = {{Ceccarelli}, C. and {Caselli}, P. and {Bockel{\'e}e-Morvan}, D. and {Mousis}, O. and {Pizzarello}, S. and {Robert}, F. and {Semenov}, D.},
        title = "{Deuterium Fractionation: The Ariadne's Thread from the Precollapse Phase to Meteorites and Comets Today}",
    booktitle = {Protostars and Planets VI},
         year = 2014,
       editor = {{Beuther}, Henrik and {Klessen}, Ralf S. and {Dullemond}, Cornelis P. and {Henning}, Thomas},
        month = jan,
        pages = {859-882},
          doi = {10.2458/azu_uapress_9780816531240-ch037},
archivePrefix = {arXiv},
       eprint = {1403.7143},
 primaryClass = {astro-ph.EP},
       adsurl = {https://ui.adsabs.harvard.edu/abs/2014prpl.conf..859C}
}

@ARTICLE{Krasnokutski2014ApPhL,
       author = {{Krasnokutski}, S.~A. and {Huisken}, F.},
        title = "{A simple and clean source of low-energy atomic carbon}",
      journal = {Appl. Phys. Lett.},
         year = 2014,
        month = sep,
       volume = {105},
       number = {11},
          eid = {113506},
        pages = {113506},
          doi = {10.1063/1.4895806},
       adsurl = {https://ui.adsabs.harvard.edu/abs/2014ApPhL.105k3506K}
}

@ARTICLE{Agundez2014A&A,
       author = {{Ag{\'u}ndez}, Marcelino and {Cernicharo}, Jos{\'e} and {Gu{\'e}lin}, Michel},
        title = "{New molecules in IRC +10216: confirmation of C$_{5}$S and tentative identification of MgCCH, NCCP, and SiH$_{3}$CN}",
      journal = {\aap},
         year = 2014,
        month = oct,
       volume = {570},
          eid = {A45},
        pages = {A45},
          doi = {10.1051/0004-6361/201424542},
archivePrefix = {arXiv},
       eprint = {1408.6306},
 primaryClass = {astro-ph.GA},
       adsurl = {https://ui.adsabs.harvard.edu/abs/2014A&A...570A..45A}
}

@ARTICLE{Bialy2015MNRAS,
       author = {{Bialy}, Shmuel and {Sternberg}, Amiel},
        title = "{CO/H$_{2}$, C/CO, OH/CO, and OH/O$_{2}$ in dense interstellar gas: from high ionization to low metallicity}",
      journal = {\mnras},
         year = 2015,
        month = jul,
       volume = {450},
       number = {4},
        pages = {4424-4445},
          doi = {10.1093/mnras/stv851},
archivePrefix = {arXiv},
       eprint = {1409.6724},
 primaryClass = {astro-ph.GA},
       adsurl = {https://ui.adsabs.harvard.edu/abs/2015MNRAS.450.4424B}
}

@ARTICLE{Fedoseev2015MNRAS,
       author = {{Fedoseev}, G. and {Ioppolo}, S. and {Zhao}, D. and {Lamberts}, T. and {Linnartz}, H.},
        title = "{Low-temperature surface formation of NH$_{3}$ and HNCO: hydrogenation of nitrogen atoms in CO-rich interstellar ice analogues}",
      journal = {\mnras},
         year = 2015,
        month = jan,
       volume = {446},
       number = {1},
        pages = {439-448},
          doi = {10.1093/mnras/stu2028},
archivePrefix = {arXiv},
       eprint = {1705.09184},
 primaryClass = {astro-ph.GA},
       adsurl = {https://ui.adsabs.harvard.edu/abs/2015MNRAS.446..439F}
}

@ARTICLE{Fedoseev2015MNRAS2,
       author = {{Fedoseev}, G. and {Cuppen}, H.~M. and {Ioppolo}, S. and {Lamberts}, T. and {Linnartz}, H.},
        title = "{Experimental evidence for glycolaldehyde and ethylene glycol formation by surface hydrogenation of CO molecules under dense molecular cloud conditions}",
      journal = {\mnras},
         year = 2015,
        month = apr,
       volume = {448},
       number = {2},
        pages = {1288-1297},
          doi = {10.1093/mnras/stu2603},
archivePrefix = {arXiv},
       eprint = {1705.09235},
 primaryClass = {astro-ph.GA},
       adsurl = {https://ui.adsabs.harvard.edu/abs/2015MNRAS.448.1288F}
}

@ARTICLE{Boorgert2015ARA&A,
       author = {{Boogert}, A.~C. Adwin and {Gerakines}, Perry A. and {Whittet}, Douglas C.~B.},
        title = "{Observations of the icy universe.}",
      journal = {\araa},
         year = 2015,
        month = aug,
       volume = {53},
        pages = {541-581},
          doi = {10.1146/annurev-astro-082214-122348},
archivePrefix = {arXiv},
       eprint = {1501.05317},
 primaryClass = {astro-ph.GA},
       adsurl = {https://ui.adsabs.harvard.edu/abs/2015ARA&A..53..541B}
}

@ARTICLE{Burton2015ApJ,
       author = {{Burton}, Michael G. and {Ashley}, Michael C.~B. and {Braiding}, Catherine and {Freeman}, Matthew and {Kulesa}, Craig and {Wolfire}, Mark G. and {Hollenbach}, David J. and {Rowell}, Gavin and {Lau}, James},
        title = "{Extended Carbon Line Emission in the Galaxy: Searching for Dark Molecular Gas along the G328 Sightline}",
      journal = {\apj},
         year = 2015,
        month = sep,
       volume = {811},
       number = {1},
          eid = {13},
        pages = {13},
          doi = {10.1088/0004-637X/811/1/13},
archivePrefix = {arXiv},
       eprint = {1508.04828},
 primaryClass = {astro-ph.GA},
       adsurl = {https://ui.adsabs.harvard.edu/abs/2015ApJ...811...13B}
}

@article{martin-domenech_uv_2015,
    title = {{UV} photoprocessing of {CO} $_{\textrm{2}}$ ice: a complete quantification of photochemistry and photon-induced desorption processes},
    volume = {584},
    issn = {0004-6361, 1432-0746},
    shorttitle = {{UV} photoprocessing of {CO} $_{\textrm{2}}$ ice},
    url = {http://www.aanda.org/10.1051/0004-6361/201526003},
    doi = {10.1051/0004-6361/201526003},
    language = {en},
    urldate = {2024-05-01},
    journal = {\aap},
    author = {Martín-Doménech, R. and Manzano-Santamaría, J. and Muñoz Caro, G. M. and Cruz-Díaz, G. A. and Chen, Y.-J. and Herrero, V. J. and Tanarro, I.},
    month = dec,
    year = {2015},
    pages = {A14},
}

@ARTICLE{Chuang2016MNRAS,
       author = {{Chuang}, K.-J. and {Fedoseev}, G. and {Ioppolo}, S. and {van Dishoeck}, E.~F. and {Linnartz}, H.},
        title = "{H-atom addition and abstraction reactions in mixed CO, H$_{2}$CO and CH$_{3}$OH ices - an extended view on complex organic molecule formation}",
      journal = {\mnras},
         year = 2016,
        month = jan,
       volume = {455},
       number = {2},
        pages = {1702-1712},
          doi = {10.1093/mnras/stv2288},
archivePrefix = {arXiv},
       eprint = {1606.01049},
 primaryClass = {astro-ph.GA},
       adsurl = {https://ui.adsabs.harvard.edu/abs/2016MNRAS.455.1702C}
}

@ARTICLE{Hudson2016PCCP,
       author = {{Hudson}, R.~L.},
        title = "{Infrared spectra and band strengths of CH3SH, an interstellar molecule}",
      journal = {Phys. Chem. Chem. Phys.},
         year = 2016,
        month = jan,
       volume = {18},
       number = {36},
        pages = {25756-25763},
          doi = {10.1039/C6CP01475E},
       adsurl = {https://ui.adsabs.harvard.edu/abs/2016PCCP...1825756H}
}

@article{chuang_h-atom_2016,
    title = {H-atom addition and abstraction reactions in mixed {CO}, {H2CO} and {CH3OH} ices – an extended view on complex organic molecule formation},
    volume = {455},
    issn = {1365-2966, 0035-8711},
    url = {http://academic.oup.com/mnras/article/455/2/1702/1106461/Hatom-addition-and-abstraction-reactions-in-mixed},
    doi = {10.1093/mnras/stv2288},
    language = {en},
    number = {2},
    urldate = {2024-10-28},
    journal = {\mnras},
    author = {Chuang, K.-J. and Fedoseev, G. and Ioppolo, S. and Van Dishoeck, E.F. and Linnartz, H.},
    month = jan,
    year = {2016},
    pages = {1702--1712},
}

@ARTICLE{Muller2016A&A,
       author = {{M{\"u}ller}, Holger S.~P. and {Belloche}, Arnaud and {Xu}, Li-Hong and {Lees}, Ronald M. and {Garrod}, Robin T. and {Walters}, Adam and {van Wijngaarden}, Jennifer and {Lewen}, Frank and {Schlemmer}, Stephan and {Menten}, Karl M.},
        title = "{Exploring molecular complexity with ALMA (EMoCA): Alkanethiols and alkanols in Sagittarius B2(N2)}",
      journal = {\aap},
         year = 2016,
        month = mar,
       volume = {587},
          eid = {A92},
        pages = {A92},
          doi = {10.1051/0004-6361/201527470},
archivePrefix = {arXiv},
       eprint = {1512.05301},
 primaryClass = {astro-ph.GA},
       adsurl = {https://ui.adsabs.harvard.edu/abs/2016A&A...587A..92M}
}

@ARTICLE{Majumdar2016MNRAS,
       author = {{Majumdar}, L. and {Gratier}, P. and {Vidal}, T. and {Wakelam}, V. and {Loison}, J.-C. and {Hickson}, K.~M. and {Caux}, E.},
        title = "{Detection of CH$_{3}$SH in protostar IRAS 16293-2422}",
      journal = {\mnras},
         year = 2016,
        month = may,
       volume = {458},
       number = {2},
        pages = {1859-1865},
          doi = {10.1093/mnras/stw457},
archivePrefix = {arXiv},
       eprint = {1602.07619},
 primaryClass = {astro-ph.GA},
       adsurl = {https://ui.adsabs.harvard.edu/abs/2016MNRAS.458.1859M}
}

@ARTICLE{Jorgensen2016A&A,
       author = {{J{\o}rgensen}, J.~K. and {van der Wiel}, M.~H.~D. and {Coutens}, A. and {Lykke}, J.~M. and {M{\"u}ller}, H.~S.~P. and {van Dishoeck}, E.~F. and {Calcutt}, H. and {Bjerkeli}, P. and {Bourke}, T.~L. and {Drozdovskaya}, M.~N. and {Favre}, C. and {Fayolle}, E.~C. and {Garrod}, R.~T. and {Jacobsen}, S.~K. and {{\"O}berg}, K.~I. and {Persson}, M.~V. and {Wampfler}, S.~F.},
        title = "{The ALMA Protostellar Interferometric Line Survey (PILS). First results from an unbiased submillimeter wavelength line survey of the Class 0 protostellar binary IRAS 16293-2422 with ALMA}",
      journal = {\aap},
         year = 2016,
        month = nov,
       volume = {595},
          eid = {A117},
        pages = {A117},
          doi = {10.1051/0004-6361/201628648},
archivePrefix = {arXiv},
       eprint = {1607.08733},
 primaryClass = {astro-ph.SR},
       adsurl = {https://ui.adsabs.harvard.edu/abs/2016A&A...595A.117J}
}

@ARTICLE{Krasnokutski2016ApJ,
       author = {{Krasnokutski}, S.~A. and {Kuhn}, M. and {Renzler}, M. and {J{\"a}ger}, C. and {Henning}, Th. and {Scheier}, P.},
        title = "{Ultra-low-temperature Reactions of Carbon Atoms with Hydrogen Molecules}",
      journal = {ApJL},
         year = 2016,
        month = feb,
       volume = {818},
       number = {2},
          eid = {L31},
        pages = {L31},
          doi = {10.3847/2041-8205/818/2/L31},
archivePrefix = {arXiv},
       eprint = {1602.09030},
 primaryClass = {astro-ph.GA},
       adsurl = {https://ui.adsabs.harvard.edu/abs/2016ApJ...818L..31K}
}

@ARTICLE{Lamberts2017JPCA,
       author = {{Lamberts}, Thanja and {K{\"a}stner}, Johannes},
        title = "{Tunneling Reaction Kinetics for the Hydrogen Abstraction Reaction H + H2S {\textrightarrow} H2 + HS in the Interstellar Medium}",
      journal = {J. Phys. Chem. A},
         year = 2017,
        month = dec,
       volume = {121},
       number = {51},
        pages = {9736-9741},
          doi = {10.1021/acs.jpca.7b10296},
archivePrefix = {arXiv},
       eprint = {1712.02088},
 primaryClass = {astro-ph.GA},
       adsurl = {https://ui.adsabs.harvard.edu/abs/2017JPCA..121.9736L}
}

@ARTICLE{Vidal2017MNRAS,
       author = {{Vidal}, Thomas H.~G. and {Loison}, Jean-Christophe and {Jaziri}, Adam Yassin and {Ruaud}, Maxime and {Gratier}, Pierre and {Wakelam}, Valentine},
        title = "{On the reservoir of sulphur in dark clouds: chemistry and elemental abundance reconciled}",
      journal = {\mnras},
         year = 2017,
        month = jul,
       volume = {469},
       number = {1},
        pages = {435-447},
          doi = {10.1093/mnras/stx828},
archivePrefix = {arXiv},
       eprint = {1704.01404},
 primaryClass = {astro-ph.GA},
       adsurl = {https://ui.adsabs.harvard.edu/abs/2017MNRAS.469..435V}
}

@ARTICLE{Albar2017NatSR,
       author = {{Albar}, J.~D. and {Summerfield}, A. and {Cheng}, T.~S. and {Davies}, A. and {Smith}, E.~F. and {Khlobystov}, A.~N. and {Mellor}, C.~J. and {Taniguchi}, T. and {Watanabe}, K. and {Foxon}, C.~T. and {Eaves}, L. and {Beton}, P.~H. and {Novikov}, S.~V.},
        title = "{An atomic carbon source for high temperature molecular beam epitaxy of graphene}",
      journal = {Sci. Rep.},
         year = 2017,
        month = jul,
       volume = {7},
          eid = {6598},
        pages = {6598},
          doi = {10.1038/s41598-017-07021-1},
       adsurl = {https://ui.adsabs.harvard.edu/abs/2017NatSR...7.6598A}
}

@ARTICLE{Oba2018NatAs,
       author = {{Oba}, Y. and {Tomaru}, T. and {Lamberts}, T. and {Kouchi}, A. and {Watanabe}, N.},
        title = "{An infrared measurement of chemical desorption from interstellar ice analogues}",
      journal = {Nat. Astron.},
         year = 2018,
        month = mar,
       volume = {2},
        pages = {228-232},
          doi = {10.1038/s41550-018-0380-9},
archivePrefix = {arXiv},
       eprint = {1810.04669},
 primaryClass = {astro-ph.GA},
       adsurl = {https://ui.adsabs.harvard.edu/abs/2018NatAs...2..228O}
}

@ARTICLE{Drozdovskaya2018MNRAS,
       author = {{Drozdovskaya}, Maria N. and {van Dishoeck}, Ewine F. and {J{\o}rgensen}, Jes K. and {Calmonte}, Ursina and {van der Wiel}, Matthijs H.~D. and {Coutens}, Audrey and {Calcutt}, Hannah and {M{\"u}ller}, Holger S.~P. and {Bjerkeli}, Per and {Persson}, Magnus V. and {Wampfler}, Susanne F. and {Altwegg}, Kathrin},
        title = "{The ALMA-PILS survey: the sulphur connection between protostars and comets: IRAS 16293-2422 B and 67P/Churyumov-Gerasimenko}",
      journal = {\mnras},
         year = 2018,
        month = jun,
       volume = {476},
       number = {4},
        pages = {4949-4964},
          doi = {10.1093/mnras/sty462},
archivePrefix = {arXiv},
       eprint = {1802.02977},
 primaryClass = {astro-ph.SR},
       adsurl = {https://ui.adsabs.harvard.edu/abs/2018MNRAS.476.4949D}
}

@ARTICLE{Lamberts2018A&A,
       author = {{Lamberts}, T.},
        title = "{From interstellar carbon monosulfide to methyl mercaptan: paths of least resistance}",
      journal = {\aap},
         year = 2018,
        month = jul,
       volume = {615},
          eid = {L2},
        pages = {L2},
          doi = {10.1051/0004-6361/201832830},
archivePrefix = {arXiv},
       eprint = {1806.02990},
 primaryClass = {astro-ph.GA},
       adsurl = {https://ui.adsabs.harvard.edu/abs/2018A&A...615L...2L}
}

@ARTICLE{Vastel2018MNRAS,
       author = {{Vastel}, Charlotte and {Qu{\'e}nard}, D. and {Le Gal}, R. and {Wakelam}, V. and {Andrianasolo}, A. and {Caselli}, P. and {Vidal}, T. and {Ceccarelli}, C. and {Lefloch}, B. and {Bachiller}, R.},
        title = "{Sulphur chemistry in the L1544 pre-stellar core}",
      journal = {\mnras},
         year = 2018,
        month = aug,
       volume = {478},
       number = {4},
        pages = {5514-5532},
          doi = {10.1093/mnras/sty1336},
archivePrefix = {arXiv},
       eprint = {1806.01102},
 primaryClass = {astro-ph.GA},
       adsurl = {https://ui.adsabs.harvard.edu/abs/2018MNRAS.478.5514V}
}

@ARTICLE{Chuang2018A&A,
       author = {{Chuang}, K.-J. and {Fedoseev}, G. and {Qasim}, D. and {Ioppolo}, S. and {van Dishoeck}, E.~F. and {Linnartz}, H.},
        title = "{H$_{2}$ chemistry in interstellar ices: the case of CO ice hydrogenation in UV irradiated CO:H$_{2}$ ice mixtures}",
      journal = {\aap},
         year = 2018,
        month = sep,
       volume = {617},
          eid = {A87},
        pages = {A87},
          doi = {10.1051/0004-6361/201833439},
archivePrefix = {arXiv},
       eprint = {1809.10507},
 primaryClass = {astro-ph.SR},
       adsurl = {https://ui.adsabs.harvard.edu/abs/2018A&A...617A..87C}
}

@ARTICLE{Hudson2018ApJ,
       author = {{Hudson}, Reggie L. and {Gerakines}, Perry A.},
        title = "{Infrared Spectra and Interstellar Sulfur: New Laboratory Results for H$_{2}$S and Four Malodorous Thiol Ices}",
      journal = {\apj},
         year = 2018,
        month = nov,
       volume = {867},
       number = {2},
          eid = {138},
        pages = {138},
          doi = {10.3847/1538-4357/aae52a},
       adsurl = {https://ui.adsabs.harvard.edu/abs/2018ApJ...867..138H}
}

@phdthesis{Chuang2018PhD,
  author = {Chuang, Kuo-Ju},
  title = {The Formation of Complex Organic Molecules in Dense Clouds: Sweet Results from the Laboratory},
  school = {Leiden University},
  address = {Leiden, The Netherlands},
  year = {2018}
}

@article{VAR018:20170145,
author = {Varandas, A. J. C. and Rocha, C. M. R.},
title = "{C$_n$($n$=2-4): current status}",
volume = {376},
year = {2018},
doi = {10.1098/rsta.2017.0145},
journal = {Phil. Trans. R. Soc. A},
pages = "20170145",
}

@ARTICLE{Muller2019A&A,
       author = {{M{\"u}ller}, Holger S.~P. and {Maeda}, Atsuko and {Thorwirth}, Sven and {Lewen}, Frank and {Schlemmer}, Stephan and {Medvedev}, Ivan R. and {Winnewisser}, Manfred and {De Lucia}, Frank C. and {Herbst}, Eric},
        title = "{Laboratory spectroscopic study of isotopic thioformaldehyde, H$_{2}$CS, and determination of its equilibrium structure}",
      journal = {\aap},
         year = 2019,
        month = jan,
       volume = {621},
          eid = {A143},
        pages = {A143},
          doi = {10.1051/0004-6361/201834517},
archivePrefix = {arXiv},
       eprint = {1812.01554},
 primaryClass = {astro-ph.SR},
       adsurl = {https://ui.adsabs.harvard.edu/abs/2019A&A...621A.143M}
}

@ARTICLE{Laas2019A&A,
       author = {{Laas}, Jacob C. and {Caselli}, Paola},
        title = "{Modeling sulfur depletion in interstellar clouds}",
      journal = {\aap},
         year = 2019,
        month = apr,
       volume = {624},
          eid = {A108},
        pages = {A108},
          doi = {10.1051/0004-6361/201834446},
archivePrefix = {arXiv},
       eprint = {1903.01232},
 primaryClass = {astro-ph.GA},
       adsurl = {https://ui.adsabs.harvard.edu/abs/2019A&A...624A.108L}
}

@ARTICLE{Bisbas2019MNRAS,
       author = {{Bisbas}, Thomas G. and {Schruba}, Andreas and {van Dishoeck}, Ewine F.},
        title = "{Simulating the atomic and molecular content of molecular clouds using probability distributions of physical parameters}",
      journal = {\mnras},
         year = 2019,
        month = may,
       volume = {485},
       number = {3},
        pages = {3097-3111},
          doi = {10.1093/mnras/stz405},
archivePrefix = {arXiv},
       eprint = {1901.11306},
 primaryClass = {astro-ph.GA},
       adsurl = {https://ui.adsabs.harvard.edu/abs/2019MNRAS.485.3097B}
}

@ARTICLE{LeGal2019ApJ,
       author = {{Le Gal}, Romane and {{\"O}berg}, Karin I. and {Loomis}, Ryan A. and {Pegues}, Jamila and {Bergner}, Jennifer B.},
        title = "{Sulfur Chemistry in Protoplanetary Disks: CS and H$_{2}$CS}",
      journal = {\apj},
         year = 2019,
        month = may,
       volume = {876},
       number = {1},
          eid = {72},
        pages = {72},
          doi = {10.3847/1538-4357/ab1416},
archivePrefix = {arXiv},
       eprint = {1903.11105},
 primaryClass = {astro-ph.GA},
       adsurl = {https://ui.adsabs.harvard.edu/abs/2019ApJ...876...72L}
}

@ARTICLE{Kama2019ApJ,
       author = {{Kama}, Mihkel and {Shorttle}, Oliver and {Jermyn}, Adam S. and {Folsom}, Colin P. and {Furuya}, Kenji and {Bergin}, Edwin A. and {Walsh}, Catherine and {Keller}, Lindsay},
        title = "{Abundant Refractory Sulfur in Protoplanetary Disks}",
      journal = {\apj},
         year = 2019,
        month = nov,
       volume = {885},
       number = {2},
          eid = {114},
        pages = {114},
          doi = {10.3847/1538-4357/ab45f8},
archivePrefix = {arXiv},
       eprint = {1908.05169},
 primaryClass = {astro-ph.EP},
       adsurl = {https://ui.adsabs.harvard.edu/abs/2019ApJ...885..114K}
}

@article{Krasnokutski_2020,
doi = {10.3847/1538-4357/ab60a1},
url = {https://doi.org/10.3847/1538-4357/ab60a1},
year = {2020},
month = {jan},
publisher = {The American Astronomical Society},
volume = {889},
number = {1},
pages = {67},
author = {Krasnokutski, Serge A. and Jäger, Cornelia and Henning, Thomas},
title = {Condensation of Atomic Carbon: Possible Routes toward Glycine},
journal = {\apj}
}

@ARTICLE{Shingledecker2020ApJ,
       author = {{Shingledecker}, Christopher N. and {Lamberts}, Thanja and {Laas}, Jacob C. and {Vasyunin}, Anton and {Herbst}, Eric and {K{\"a}stner}, Johannes and {Caselli}, Paola},
        title = "{Efficient Production of S$_{8}$ in Interstellar Ices: The Effects of Cosmic-Ray-driven Radiation Chemistry and Nondiffusive Bulk Reactions}",
      journal = {\apj},
         year = 2020,
        month = jan,
       volume = {888},
       number = {1},
          eid = {52},
        pages = {52},
          doi = {10.3847/1538-4357/ab5360},
archivePrefix = {arXiv},
       eprint = {1911.01239},
 primaryClass = {astro-ph.GA},
       adsurl = {https://ui.adsabs.harvard.edu/abs/2020ApJ...888...52S}
}

@ARTICLE{Codella2020A&A,
       author = {{Codella}, C. and {Podio}, L. and {Garufi}, A. and {Perrero}, J. and {Ugliengo}, P. and {Fedele}, D. and {Favre}, C. and {Bianchi}, E. and {Ceccarelli}, C. and {Mercimek}, S. and {Bacciotti}, F. and {Rygl}, K.~L.~J. and {Testi}, L.},
        title = "{ALMA chemical survey of disk-outflow sources in Taurus (ALMA-DOT). IV. Thioformaldehyde (H$_{2}$CS) in protoplanetary discs: spatial distributions and binding energies}",
      journal = {\aap},
         year = 2020,
        month = dec,
       volume = {644},
          eid = {A120},
        pages = {A120},
          doi = {10.1051/0004-6361/202039309},
archivePrefix = {arXiv},
       eprint = {2011.02305},
 primaryClass = {astro-ph.EP},
       adsurl = {https://ui.adsabs.harvard.edu/abs/2020A&A...644A.120C}
}

@article{chuang_formation_2020,
    title = {Formation of complex molecules in translucent clouds: acetaldehyde, vinyl alcohol, ketene, and ethanol via “nonenergetic” processing of {C} $_{\textrm{2}}$ {H} $_{\textrm{2}}$ ice},
    volume = {635},
    copyright = {http://creativecommons.org/licenses/by/4.0},
    issn = {0004-6361, 1432-0746},
    shorttitle = {Formation of complex molecules in translucent clouds},
    url = {https://www.aanda.org/10.1051/0004-6361/201937302},
    doi = {10.1051/0004-6361/201937302},
    language = {en},
    urldate = {2024-09-05},
    journal = {\aap},
    author = {Chuang, K.-J. and Fedoseev, G. and Qasim, D. and Ioppolo, S. and Jäger, C. and Henning, Th. and Palumbo, M. E. and Van Dishoeck, E. F. and Linnartz, H.},
    month = mar,
    year = {2020},
    pages = {A199},
}

@ARTICLE{Qasim2020NatAs,
       author = {{Qasim}, D. and {Fedoseev}, G. and {Chuang}, K.-J. and {He}, J. and {Ioppolo}, S. and {van Dishoeck}, E.~F. and {Linnartz}, H.},
        title = "{An experimental study of the surface formation of methane in interstellar molecular clouds}",
      journal = {Nat. Astron.},
         year = 2020,
        month = apr,
       volume = {4},
        pages = {781-785},
          doi = {10.1038/s41550-020-1054-y},
archivePrefix = {arXiv},
       eprint = {2004.02506},
 primaryClass = {astro-ph.SR},
       adsurl = {https://ui.adsabs.harvard.edu/abs/2020NatAs...4..781Q}
}

@ARTICLE{Simoncic2020A&A,
       author = {{Simon{\v{c}}i{\v{c}}}, M. and {Semenov}, D. and {Krasnokutski}, S. and {Henning}, Th. and {J{\"a}ger}, C.},
        title = "{Sensitivity of gas-grain chemical models to surface reaction barriers. Effect from a key carbon-insertion reaction, C + H$_{2}$ {\textrightarrow} CH$_{2}$}",
      journal = {\aap},
         year = 2020,
        month = may,
       volume = {637},
          eid = {A72},
        pages = {A72},
          doi = {10.1051/0004-6361/202037648},
archivePrefix = {arXiv},
       eprint = {2003.14129},
 primaryClass = {astro-ph.GA},
       adsurl = {https://ui.adsabs.harvard.edu/abs/2020A&A...637A..72S}
}

@ARTICLE{Qasim2020RScI,
       author = {{Qasim}, D. and {Witlox}, M.~J.~A. and {Fedoseev}, G. and {Chuang}, K.-J. and {Banu}, T. and {Krasnokutski}, S.~A. and {Ioppolo}, S. and {K{\"a}stner}, J. and {van Dishoeck}, E.~F. and {Linnartz}, H.},
        title = "{A cryogenic ice setup to simulate carbon atom reactions in interstellar ices}",
      journal = {Rev. Sci. Instrum.},
         year = 2020,
        month = may,
       volume = {91},
       number = {5},
          eid = {054501},
        pages = {054501},
          doi = {10.1063/5.0003692},
archivePrefix = {arXiv},
       eprint = {2004.06681},
 primaryClass = {astro-ph.SR},
       adsurl = {https://ui.adsabs.harvard.edu/abs/2020RScI...91e4501Q}
}

@ARTICLE{Jin2020ApJS,
       author = {{Jin}, Mihwa and {Garrod}, Robin T.},
        title = "{Formation of Complex Organic Molecules in Cold Interstellar Environments through Nondiffusive Grain-surface and Ice-mantle Chemistry}",
      journal = {\apjs},
         year = 2020,
        month = aug,
       volume = {249},
       number = {2},
          eid = {26},
        pages = {26},
          doi = {10.3847/1538-4365/ab9ec8},
archivePrefix = {arXiv},
       eprint = {2006.11127},
 primaryClass = {astro-ph.GA},
       adsurl = {https://ui.adsabs.harvard.edu/abs/2020ApJS..249...26J}
}

@misc{MOLPRO,
title="{{MOLPRO}, version 2020.1, A Package of \emph{Ab Initio} Programs}",
author={Werner, H.-J. and Knowles, P. J. and Knizia, G. and Manby, F. R. and Sch\"{u}tz, M. and others},
note={see:~http://www.molpro.net},
year={2020}
}

@article{oberg_astrochemistry_2021,
    title = {Astrochemistry and compositions of planetary systems},
    volume = {893},
    issn = {03701573},
    url = {https://linkinghub.elsevier.com/retrieve/pii/S0370157320303446},
    doi = {10.1016/j.physrep.2020.09.004},
    language = {en},
    urldate = {2024-08-02},
    journal = {Phys. Rep.},
    author = {Öberg, Karin I. and Bergin, Edwin A.},
    month = jan,
    year = {2021},
    pages = {1--48},
}

@ARTICLE{Cernicharo2021A&A,
       author = {{Cernicharo}, J. and {Cabezas}, C. and {Ag{\'u}ndez}, M. and {Tercero}, B. and {Pardo}, J.~R. and {Marcelino}, N. and {Gallego}, J.~D. and {Tercero}, F. and {L{\'o}pez-P{\'e}rez}, J.~A. and {de Vicente}, P.},
        title = "{TMC-1, the starless core sulfur factory: Discovery of NCS, HCCS, H$_{2}$CCS, H$_{2}$CCCS, and C$_{4}$S and detection of C$_{5}$S}",
      journal = {\aap},
         year = 2021,
        month = apr,
       volume = {648},
          eid = {L3},
        pages = {L3},
          doi = {10.1051/0004-6361/202140642},
archivePrefix = {arXiv},
       eprint = {2103.12431},
 primaryClass = {astro-ph.GA},
       adsurl = {https://ui.adsabs.harvard.edu/abs/2021A&A...648L...3C}
}

@ARTICLE{EnriqueRomero2021A&A,
       author = {{Enrique-Romero}, Joan and {Ceccarelli}, Cecilia and {Rimola}, Albert and {Skouteris}, Dimitrios and {Balucani}, Nadia and {Ugliengo}, Piero},
        title = "{Theoretical computations on the efficiency of acetaldehyde formation on interstellar icy grains}",
      journal = {\aap},
         year = 2021,
        month = nov,
       volume = {655},
          eid = {A9},
        pages = {A9},
          doi = {10.1051/0004-6361/202141531},
archivePrefix = {arXiv},
       eprint = {2108.12410},
 primaryClass = {astro-ph.GA},
       adsurl = {https://ui.adsabs.harvard.edu/abs/2021A&A...655A...9E}
}

@article{Molpeceres2021JPCL,
  title        = {Carbon Atom Reactivity with Amorphous Solid Water: H_{2}O‐Catalyzed Formation of H_{2}CO},
  author       = {Molpeceres, Germ{\'a}n and K{\"a}stner, Johannes and Fedoseev, Gleb and Qasim, Danna and Sch{\"o}mig, Richard and Linnartz, Harold and Lamberts, Thanja},
  journal      = {J. Phys. Chem. Lett.},
  year         = {2021},
  volume       = {12},
  number       = {44},
  pages        = {10854--10860},
  doi          = {10.1021/acs.jpclett.1c02760},
  note         = {Published online November 2 2021},
}

@article{RodriguezAlmeida2021ApJL,
  author  = {Rodr{\'i}guez-Almeida, L. F. and Rivilla, V. M. and Jim{\'e}nez-Serra, I. and Melosso, M. and Colzi, L. and Zeng, S. and Tercero, B. and de Vicente, P. and Mart{\'i}n, S. and Requena-Torres, M. A. and Mart{\'i}n-Pintado, J. and Rico-Villas, F.},
  title   = {Thiols in the Interstellar Medium: First Detection of HC(O)SH and Confirmation of C$_2$H$_5$SH},
  journal = {ApJL},
  year    = {2021},
  volume  = {912},
  number  = {1},
  pages   = {L11},
  doi     = {10.3847/2041-8213/abf7cb}
}

@article{cazaux_photoprocessing_2022,
    title = {Photoprocessing of {H} $_{\textrm{2}}$ {S} on dust grains: {Building} {S} chains in translucent clouds and comets},
    volume = {657},
    copyright = {https://www.edpsciences.org/en/authors/copyright-and-licensing},
    issn = {0004-6361, 1432-0746},
    shorttitle = {Photoprocessing of {H} $_{\textrm{2}}$ {S} on dust grains},
    url = {https://www.aanda.org/10.1051/0004-6361/202141861},
    doi = {10.1051/0004-6361/202141861},
    language = {en},
    urldate = {2024-05-01},
    journal = {\aap},
    author = {Cazaux, S. and Carrascosa, H. and Muñoz Caro, G. M. and Caselli, P. and Fuente, A. and Navarro-Almaida, D. and Riviére-Marichalar, P.},
    month = jan,
    year = {2022},
    pages = {A100},
}

@article{fedoseev_hydrogenation_2022,
    title = {Hydrogenation of {Accreting} {C} {Atoms} and {CO} {Molecules}–{Simulating} {Ketene} and {Acetaldehyde} {Formation} {Under} {Dark} and {Translucent} {Cloud} {Conditions}},
    volume = {924},
    issn = {0004-637X, 1538-4357},
    url = {https://iopscience.iop.org/article/10.3847/1538-4357/ac3834},
    doi = {10.3847/1538-4357/ac3834},
    language = {en},
    number = {2},
    urldate = {2025-08-15},
    journal = {\apj},
    author = {Fedoseev, Gleb and Qasim, Danna and Chuang, Ko-Ju and Ioppolo, Sergio and Lamberts, Thanja and Van Dishoeck, Ewine F. and Linnartz, Harold},
    month = jan,
    year = {2022},
    pages = {110},
}

@ARTICLE{Garrod2022ApJ,
       author = {{Garrod}, Robin T. and {Jin}, Miwha and {Matis}, Kayla A. and {Jones}, Dylan and {Willis}, Eric R. and {Herbst}, Eric},
        title = "{Formation of Complex Organic Molecules in Hot Molecular Cores through Nondiffusive Grain-surface and Ice-mantle Chemistry}",
      journal = {\apjs},
         year = 2022,
        month = mar,
       volume = {259},
       number = {1},
          eid = {1},
        pages = {1},
          doi = {10.3847/1538-4365/ac3131},
archivePrefix = {arXiv},
       eprint = {2110.09743},
 primaryClass = {astro-ph.GA},
       adsurl = {https://ui.adsabs.harvard.edu/abs/2022ApJS..259....1G}
}

@ARTICLE{Lamberts2022ApJ,
       author = {{Lamberts}, Thanja and {Fedoseev}, Gleb and {van Hemert}, Marc C. and {Qasim}, Danna and {Chuang}, Ko-Ju and {Santos}, Julia C. and {Linnartz}, Harold},
        title = "{Methane Formation in Cold Regions from Carbon Atoms and Molecular Hydrogen}",
      journal = {\apj},
         year = 2022,
        month = mar,
       volume = {928},
       number = {1},
          eid = {48},
        pages = {48},
          doi = {10.3847/1538-4357/ac51d1},
archivePrefix = {arXiv},
       eprint = {2110.15881},
 primaryClass = {astro-ph.GA},
       adsurl = {https://ui.adsabs.harvard.edu/abs/2022ApJ...928...48L}
}

@ARTICLE{Caselli2022ApJ,
       author = {{Caselli}, Paola and {Pineda}, Jaime E. and {Sipil{\"a}}, Olli and {Zhao}, Bo and {Redaelli}, Elena and {Spezzano}, Silvia and {Maureira}, Maria Jos{\'e} and {Alves}, Felipe and {Bizzocchi}, Luca and {Bourke}, Tyler L. and {Chac{\'o}n-Tanarro}, Ana and {Friesen}, Rachel and {Galli}, Daniele and {Harju}, Jorma and {Jim{\'e}nez-Serra}, Izaskun and {Keto}, Eric and {Li}, Zhi-Yun and {Padovani}, Marco and {Schmiedeke}, Anika and {Tafalla}, Mario and {Vastel}, Charlotte},
        title = "{The Central 1000 au of a Prestellar Core Revealed with ALMA. II. Almost Complete Freeze-out}",
      journal = {\apj},
         year = 2022,
        month = apr,
       volume = {929},
       number = {1},
          eid = {13},
        pages = {13},
          doi = {10.3847/1538-4357/ac5913},
archivePrefix = {arXiv},
       eprint = {2202.13374},
 primaryClass = {astro-ph.SR},
       adsurl = {https://ui.adsabs.harvard.edu/abs/2022ApJ...929...13C}
}

@ARTICLE{Altwegg2022MNRAS,
       author = {{Altwegg}, K. and {Combi}, M. and {Fuselier}, S.~A. and {H{\"a}nni}, N. and {De Keyser}, J. and {Mahjoub}, A. and {M{\"u}ller}, D.~R. and {Pestoni}, B. and {Rubin}, M. and {Wampfler}, S.~F.},
        title = "{Abundant ammonium hydrosulphide embedded in cometary dust grains}",
      journal = {\mnras},
         year = 2022,
        month = nov,
       volume = {516},
       number = {3},
        pages = {3900-3910},
          doi = {10.1093/mnras/stac2440},
archivePrefix = {arXiv},
       eprint = {2208.11396},
 primaryClass = {astro-ph.EP},
       adsurl = {https://ui.adsabs.harvard.edu/abs/2022MNRAS.516.3900A}
}

@ARTICLE{Boogert2022ApJ,
       author = {{Boogert}, A.~C.~A. and {Brewer}, K. and {Brittain}, A. and {Emerson}, K.~S.},
        title = "{Survey of Ices toward Massive Young Stellar Objects. I. OCS, CO, OCN$^{-}$, and CH$_{3}$OH}",
      journal = {\apj},
         year = 2022,
        month = dec,
       volume = {941},
       number = {1},
          eid = {32},
        pages = {32},
          doi = {10.3847/1538-4357/ac9b4a},
archivePrefix = {arXiv},
       eprint = {2210.12639},
 primaryClass = {astro-ph.GA},
       adsurl = {https://ui.adsabs.harvard.edu/abs/2022ApJ...941...32B}
}

@ARTICLE{Yang2022ApJ,
       author = {{Yang}, Yao-Lun and {Green}, Joel D. and {Pontoppidan}, Klaus M. and {Bergner}, Jennifer B. and {Cleeves}, L. Ilsedore and {Evans}, II, Neal J. and {Garrod}, Robin T. and {Jin}, Miwha and {Kim}, Chul Hwan and {Kim}, Jaeyeong and {Lee}, Jeong-Eun and {Sakai}, Nami and {Shingledecker}, Christopher N. and {Shope}, Brielle and {Tobin}, John J. and {van Dishoeck}, Ewine F.},
        title = "{CORINOS. I. JWST/MIRI Spectroscopy and Imaging of a Class 0 Protostar IRAS 15398{\textendash}3359}",
      journal = {ApJL},
         year = 2022,
        month = dec,
       volume = {941},
       number = {1},
          eid = {L13},
        pages = {L13},
          doi = {10.3847/2041-8213/aca289},
archivePrefix = {arXiv},
       eprint = {2208.10673},
 primaryClass = {astro-ph.SR},
       adsurl = {https://ui.adsabs.harvard.edu/abs/2022ApJ...941L..13Y}
}

@ARTICLE{Fuente2023A&A,
       author = {{Fuente}, A. and {Rivi{\`e}re-Marichalar}, P. and {Beitia-Antero}, L. and {Caselli}, P. and {Wakelam}, V. and {Esplugues}, G. and {Rodr{\'\i}guez-Baras}, M. and {Navarro-Almaida}, D. and {Gerin}, M. and {Kramer}, C. and {Bachiller}, R. and {Goicoechea}, J.~R. and {Jim{\'e}nez-Serra}, I. and {Loison}, J.~C. and {Ivlev}, A. and {Mart{\'\i}n-Dom{\'e}nech}, R. and {Spezzano}, S. and {Roncero}, O. and {Mu{\~n}oz-Caro}, G. and {Cazaux}, S. and {Marcelino}, N.},
        title = "{Gas phase Elemental abundances in Molecular cloudS (GEMS). VII. Sulfur elemental abundance}",
      journal = {\aap},
         year = 2023,
        month = feb,
       volume = {670},
          eid = {A114},
        pages = {A114},
          doi = {10.1051/0004-6361/202244843},
archivePrefix = {arXiv},
       eprint = {2212.03742},
 primaryClass = {astro-ph.GA},
       adsurl = {https://ui.adsabs.harvard.edu/abs/2023A&A...670A.114F}
}

@ARTICLE{McClure2023NatAs,
       author = {{McClure}, M.~K. and {Rocha}, W.~R.~M. and {Pontoppidan}, K.~M. and {Crouzet}, N. and {Chu}, L.~E.~U. and {Dartois}, E. and {Lamberts}, T. and {Noble}, J.~A. and {Pendleton}, Y.~J. and {Perotti}, G. and {Qasim}, D. and {Rachid}, M.~G. and {Smith}, Z.~L. and {Sun}, Fengwu and {Beck}, Tracy L. and {Boogert}, A.~C.~A. and {Brown}, W.~A. and {Caselli}, P. and {Charnley}, S.~B. and {Cuppen}, Herma M. and {Dickinson}, H. and {Drozdovskaya}, M.~N. and {Egami}, E. and {Erkal}, J. and {Fraser}, H. and {Garrod}, R.~T. and {Harsono}, D. and {Ioppolo}, S. and {Jim{\'e}nez-Serra}, I. and {Jin}, M. and {J{\o}rgensen}, J.~K. and {Kristensen}, L.~E. and {Lis}, D.~C. and {McCoustra}, M.~R.~S. and {McGuire}, Brett A. and {Melnick}, G.~J. and {{\"O}berg}, Karin I. and {Palumbo}, M.~E. and {Shimonishi}, T. and {Sturm}, J.~A. and {van Dishoeck}, E.~F. and {Linnartz}, H.},
        title = "{An Ice Age JWST inventory of dense molecular cloud ices}",
      journal = {Nat. Astron.},
         year = 2023,
        month = apr,
       volume = {7},
        pages = {431-443},
          doi = {10.1038/s41550-022-01875-w},
archivePrefix = {arXiv},
       eprint = {2301.09140},
 primaryClass = {astro-ph.GA},
       adsurl = {https://ui.adsabs.harvard.edu/abs/2023NatAs...7..431M}
}

@ARTICLE{Guo2023RJPCA,
       author = {{Guo}, Tian and {Xu}, Jiangang and {Li}, Linna and {Cao}, Zhanli and {Zhang}, Yunguang},
        title = "{Ab Initio Potential Energy Surfaces and Vibrational Spectra of Thioformaldehyde}",
      journal = {Russ. J. Phys. Chem. A},
         year = 2023,
        month = sep,
       volume = {97},
       number = {9},
        pages = {1953-1963},
          doi = {10.1134/S0036024423090248},
       adsurl = {https://ui.adsabs.harvard.edu/abs/2023RJPCA..97.1953G}
}

@ARTICLE{Santos2023A&A,
       author = {{Santos}, J.~C. and {Linnartz}, H. and {Chuang}, K.-J.},
        title = "{Interaction of H$_{2}$S with H atoms on grain surfaces under molecular cloud conditions}",
      journal = {\aap},
         year = 2023,
        month = oct,
       volume = {678},
          eid = {A112},
        pages = {A112},
          doi = {10.1051/0004-6361/202347348},
archivePrefix = {arXiv},
       eprint = {2308.07960},
 primaryClass = {astro-ph.GA},
       adsurl = {https://ui.adsabs.harvard.edu/abs/2023A&A...678A.112S}
}

@ARTICLE{Tsuge2023NatAs,
       author = {{Tsuge}, Masashi and {Molpeceres}, Germ{\'a}n and {Aikawa}, Yuri and {Watanabe}, Naoki},
        title = "{Surface diffusion of carbon atoms as a driver of interstellar organic chemistry}",
      journal = {Nat. Astron.},
         year = 2023,
        month = nov,
       volume = {7},
        pages = {1351-1358},
          doi = {10.1038/s41550-023-02071-0},
archivePrefix = {arXiv},
       eprint = {2308.02181},
 primaryClass = {astro-ph.GA},
       adsurl = {https://ui.adsabs.harvard.edu/abs/2023NatAs...7.1351T}
}

@ARTICLE{Shao2023JPCL,
       author = {{Shao}, Xin and {Wu}, Zhuang and {Wang}, Lina and {Rauhut}, Guntram and {Zeng}, Xiaoqing},
        title = "{Identification and Photochemistry of the Mercaptomethyl Radical}",
      journal = {J. Phys. Chem. Lett.},
         year = 2023,
        month = nov,
       volume = {14},
       number = {46},
        pages = {10450-10456},
          doi = {10.1021/acs.jpclett.3c02526},
       adsurl = {https://ui.adsabs.harvard.edu/abs/2023JPCL...1410450S}
}

@ARTICLE{Ferrero2024ApJ,
       author = {{Ferrero}, Stefano and {Ceccarelli}, Cecilia and {Ugliengo}, Piero and {Sodupe}, Mariona and {Rimola}, Albert},
        title = "{Formation of Interstellar Complex Organic Molecules on Water-rich Ices Triggered by Atomic Carbon Freezing}",
      journal = {\apj},
         year = 2024,
        month = jan,
       volume = {960},
       number = {1},
          eid = {22},
        pages = {22},
          doi = {10.3847/1538-4357/ad0547},
archivePrefix = {arXiv},
       eprint = {2310.14831},
 primaryClass = {astro-ph.GA},
       adsurl = {https://ui.adsabs.harvard.edu/abs/2024ApJ...960...22F}
}

@ARTICLE{Martin-Domenech2024ApJ,
       author = {{Mart{\'\i}n-Dom{\'e}nech}, Rafael and {DelFranco}, Alexander and {{\"O}berg}, Karin I. and {Rajappan}, Mahesh},
        title = "{Organic Chemistry in the H$_{2}$-bearing, CO-rich Interstellar Ice Layer at Temperatures Relevant to Dense Cloud Interiors}",
      journal = {\apj},
         year = 2024,
        month = feb,
       volume = {962},
       number = {2},
          eid = {107},
        pages = {107},
          doi = {10.3847/1538-4357/ad187e},
archivePrefix = {arXiv},
       eprint = {2401.04024},
 primaryClass = {astro-ph.GA},
       adsurl = {https://ui.adsabs.harvard.edu/abs/2024ApJ...962..107M}
}

@ARTICLE{Cuppen2024ARA&A,
       author = {{Cuppen}, Herma M. and {Linnartz}, H. and {Ioppolo}, S.},
        title = "{Laboratory and Computational Studies of Interstellar Ices}",
      journal = {\araa},
         year = 2024,
        month = sep,
       volume = {62},
       number = {1},
        pages = {243-286},
          doi = {10.1146/annurev-astro-071221-052732},
archivePrefix = {arXiv},
       eprint = {2407.06657},
 primaryClass = {astro-ph.GA},
       adsurl = {https://ui.adsabs.harvard.edu/abs/2024ARA&A..62..243C}
}

@ARTICLE{Vitorino2024MNRAS,
       author = {{Vitorino}, J. and {Loison}, J.-C. and {Wakelam}, V. and {Congiu}, E. and {Dulieu}, F.},
        title = "{Sulphur storage in cold molecular clouds: the case of the NH$_{4}$$^{+}$SH$^{-}$ salt on interstellar dust grains}",
      journal = {\mnras},
         year = 2024,
        month = sep,
       volume = {533},
       number = {1},
        pages = {52-62},
          doi = {10.1093/mnras/stae1747},
       adsurl = {https://ui.adsabs.harvard.edu/abs/2024MNRAS.533...52V}
}

@article{Carrascosa2024SulphurChains,
  author  = {Carrascosa, H. and Mu{\~n}oz Caro, G. M. and Mart{\'i}n-Dom{\'e}nech, R. and Cazaux, S. and Chen, Y.-J. and Fuente, A.},
  title   = {Formation and desorption of sulphur chains ({H$_2$S$_x$} and {S$_x$}) in cometary ice: effects of ice composition and temperature},
  journal = {\mnras},
  year    = {2024},
  volume  = {533},
  number  = {1},
  pages   = {967--978},
  doi     = {10.1093/mnras/stae1768}
}

@ARTICLE{Slavicinska2025A&A,
       author = {{Slavicinska}, K. and {Boogert}, A.~C.~A. and {Tychoniec}, {\L}. and {van Dishoeck}, E.~F. and {van Gelder}, M.~L. and {Navarro}, M.~G. and {Santos}, J.~C. and {Klaassen}, P.~D. and {Kavanagh}, P.~J. and {Chuang}, K.-J.},
        title = "{Ammonium hydrosulfide (NH$_{4}$SH) as a potentially significant sulfur sink in interstellar ices}",
      journal = {\aap},
         year = 2025,
        month = jan,
       volume = {693},
          eid = {A146},
        pages = {A146},
          doi = {10.1051/0004-6361/202451383},
archivePrefix = {arXiv},
       eprint = {2410.02860},
 primaryClass = {astro-ph.GA},
       adsurl = {https://ui.adsabs.harvard.edu/abs/2025A&A...693A.146S}
}

@ARTICLE{Bunn2025ApJ,
       author = {{Bunn}, Hayley A. and {Spezzano}, Silvia and {Coudert}, Laurent H. and {Guillemin}, Jean-Claude and {Lin}, Yuxin and {Endres}, Christian P. and {Billinghurst}, Brant and {Pirali}, Olivier and {J{\o}rgensen}, Jes and {Lattanzi}, Valerio and {Caselli}, Paola},
        title = "{Laboratory Rotational Spectroscopy Leads to the First Interstellar Detection of Singly Deuterated Methyl Mercaptan (CH$_{2}$DSH)}",
      journal = {ApJL},
         year = 2025,
        month = feb,
       volume = {980},
       number = {1},
          eid = {L13},
        pages = {L13},
          doi = {10.3847/2041-8213/adacea},
archivePrefix = {arXiv},
       eprint = {2501.15944},
 primaryClass = {astro-ph.GA},
       adsurl = {https://ui.adsabs.harvard.edu/abs/2025ApJ...980L..13B}
}

@ARTICLE{Narayanan2025ApJ,
       author = {{Narayanan}, Suchitra and {Piacentino}, Elettra L. and {{\"O}berg}, Karin I. and {Rajappan}, Mahesh},
        title = "{Thermal Desorption Kinetics, Binding Energies, and Entrapment of Methyl Mercaptan Ices}",
      journal = {\apj},
         year = 2025,
        month = jun,
       volume = {986},
       number = {1},
          eid = {10},
        pages = {10},
          doi = {10.3847/1538-4357/adc7b6},
archivePrefix = {arXiv},
       eprint = {2504.01102},
 primaryClass = {astro-ph.EP},
       adsurl = {https://ui.adsabs.harvard.edu/abs/2025ApJ...986...10N}
}

@ARTICLE{Herath2025NatCo,
       author = {{Herath}, Ashanie and {McAnally}, Mason and {Turner}, Andrew M. and {Wang}, Jia and {Marks}, Joshua H. and {Fortenberry}, Ryan C. and {Garcia-Alvarez}, Jorge C. and {Gozem}, Samer and {Kaiser}, Ralf I.},
        title = "{Missing interstellar sulfur in inventories of polysulfanes and molecular octasulfur crowns}",
      journal = {Nat. Commun.},
         year = 2025,
        month = jul,
       volume = {16},
       number = {1},
          eid = {5571},
        pages = {5571},
          doi = {10.1038/s41467-025-61259-2},
       adsurl = {https://ui.adsabs.harvard.edu/abs/2025NatCo..16.5571H}
}

@ARTICLE{Borshcheva2025ApJ,
       author = {{Borshcheva}, Katerina and {Fedoseev}, Gleb and {Punanova}, Anna F. and {Caselli}, Paola and {Jim{\'e}nez-Serra}, Izaskun and {Vasyunin}, Anton I.},
        title = "{Formation of Complex Organic Molecules in Prestellar Cores: The Role of Nondiffusive Grain Chemistry}",
      journal = {\apj},
         year = 2025,
        month = sep,
       volume = {990},
       number = {2},
          eid = {163},
        pages = {163},
          doi = {10.3847/1538-4357/adea73},
archivePrefix = {arXiv},
       eprint = {2507.00338},
 primaryClass = {astro-ph.GA},
       adsurl = {https://ui.adsabs.harvard.edu/abs/2025ApJ...990..163B}
}

@article{Agundez2025A&A,
  author  = {Ag{\'u}ndez, M. and Molpeceres, G. and Cabezas, C. and Marcelino, N. and Tercero, B. and Fuentetaja, R. and de Vicente, P. and Cernicharo, J.},
  title   = {Detection of thioacetaldehyde (CH$_3$CHS) in TMC-1: Sulfur--oxygen differentiation along the hydrogenation sequence},
  journal = {\aap},
  year    = {2025},
  volume  = {693},
  pages   = {L20},
  doi     = {10.1051/0004-6361/202453459}
}

@article{SanzNovo2025ApJL,
  author  = {Sanz-Novo, Miguel and Rivilla, V{\'i}ctor M. and Endres, Christian P. and Lattanzi, Valerio and Jim{\'e}nez-Serra, Izaskun and Colzi, Laura and Zeng, Shaoshan and Meg{\'i}as, Andr{\'e}s and L{\'o}pez-Gallifa, {\'A}lvaro and Mart{\'i}nez-Henares, Antonio and San Andr{\'e}s, David and Tercero, Bel{\'e}n and de Vicente, Pablo and Mart{\'i}n, Sergio and Requena-Torres, Miguel A. and Caselli, Paola and Mart{\'i}n-Pintado, Jes{\'u}s},
  title   = {On the Abiotic Origin of Dimethyl Sulfide: Discovery of Dimethyl Sulfide in the Interstellar Medium},
  journal = {ApJL},
  year    = {2025},
  volume  = {980},
  number  = {2},
  pages   = {L37},
  doi     = {10.3847/2041-8213/adafa7}
}

@ARTICLE{Martinez-Bachs2026ApJ,
       author = {{Mart{\'\i}nez-Bachs}, Berta and {Rimola}, Albert},
        title = "{Nondiffusive Formation of Acetaldehyde on Interstellar Ices: An Atomistic Perspective}",
      journal = {ApJL},
         year = 2026,
        month = apr,
       volume = {1000},
       number = {2},
          eid = {L58},
        pages = {L58},
          doi = {10.3847/2041-8213/ae4a8e},
       adsurl = {https://ui.adsabs.harvard.edu/abs/2026ApJ..1000L..58M}
}

@ARTICLE{DiGenova2026A&A,
       author = {{Di Genova}, Gabriella and {Pareras}, Gerard and {Dominguez-Dalmases}, Christian and {Balucani}, Nadia and {Ceccarelli}, Cecilia and {Rimola}, Albert},
        title = "{Quantum mechanical atomistic investigation of the hydrogen addition to carbon monosulfide on interstellar water ice}",
      journal = {\aap},
         year = 2026,
        month = jul,
       volume = {711},
          eid = {A269},
        pages = {A269},
          doi = {10.1051/0004-6361/202660371},
       adsurl = {https://ui.adsabs.harvard.edu/abs/2026A&A...711A.269D}
}

@article{Araki2026NatAs,
  author  = {Araki, Mitsunori and Sanz-Novo, Miguel and Endres, Christian P. and Caselli, Paola and Rivilla, V{\'i}ctor M. and Jim{\'e}nez-Serra, Izaskun and Colzi, Laura and Zeng, Shaoshan and Meg{\'i}as, Andr{\'e}s and L{\'o}pez-Gallifa, {\'A}lvaro and Mart{\'i}nez-Henares, Antonio and San Andr{\'e}s, David and Mart{\'i}n, Sergio and Requena-Torres, Miguel A. and Garc{\'i}a de la Concepci{\'o}n, Juan and Lattanzi, Valerio},
  title   = {A detection of sulfur-bearing cyclic hydrocarbons in space},
  journal = {Nat. Astron.},
  year    = {2026},
  volume  = {10},
  pages   = {401--409},
  doi     = {10.1038/s41550-025-02749-7}
}

@ARTICLE{Nguyen2026MNRAS,
       author = {{Nguyen}, Thanh and {Molpeceres}, Germ{\'a}n and {Oba}, Yasuhiro and {Ag{\'u}ndez}, Marcelino and {Esplugues}, Gisela and {Cernicharo}, Jos{\'e} and {Watanabe}, Naoki},
        title = "{Experimental and computational studies of the hydrogenation of carbon disulfide (CS$_{2}$) on ice analogues}",
      journal = {\mnras},
         year = 2026,
        month = apr,
       volume = {547},
       number = {3},
          eid = {stag396},
        pages = {stag396},
          doi = {10.1093/mnras/stag396},
       adsurl = {https://ui.adsabs.harvard.edu/abs/2026MNRAS.547ag396N}
}

\begin{appendix}

\section{\texorpdfstring{Monomeric H$_2$CS desorption signals}{Monomeric H2CS desorption signals}}
\label{app:h2cs-monomer-desorption}

\begin{figure}[!htbp]
    \centering    \includegraphics[width=\columnwidth]{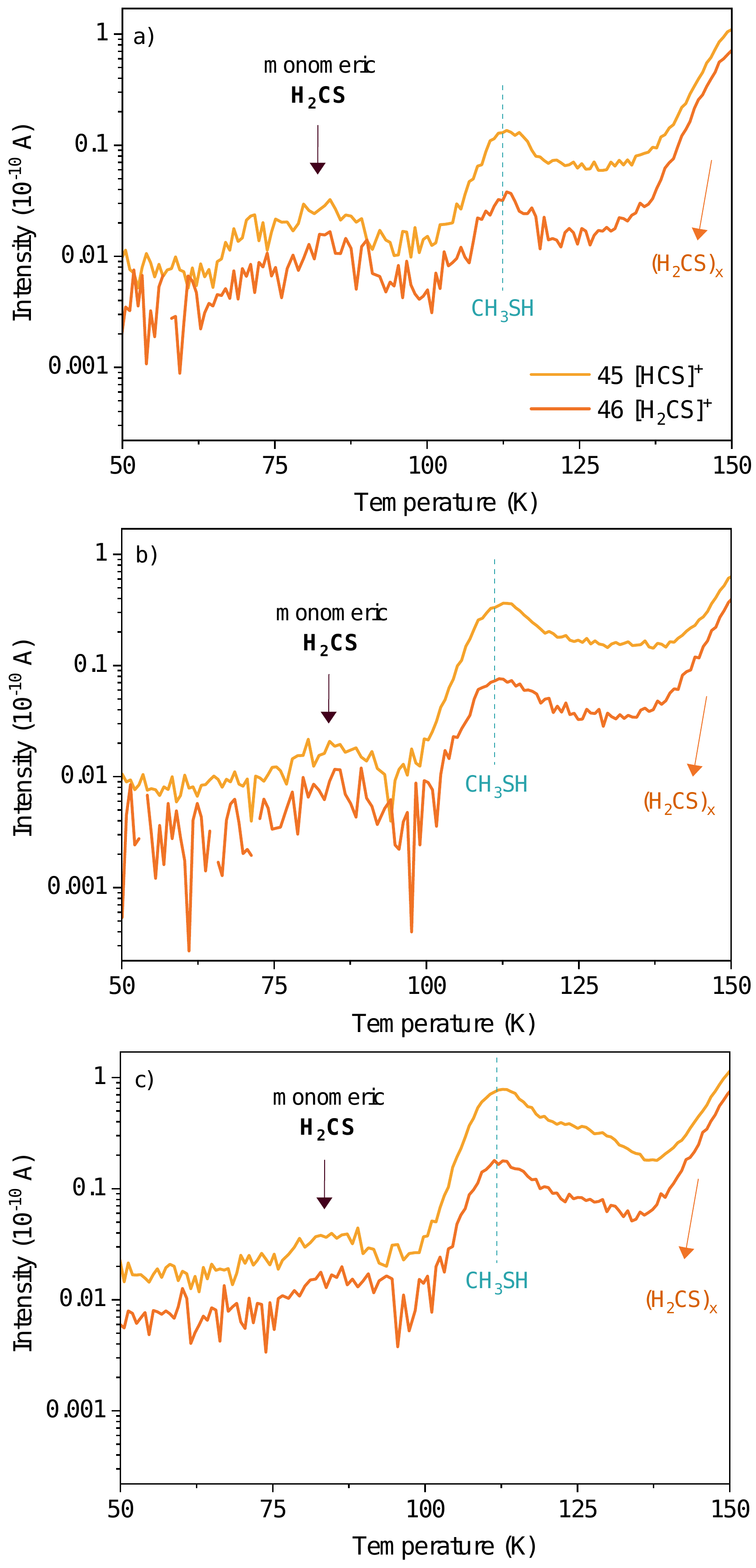}
    \caption{Enlarged view of the TPD-QMS data corresponding to panels (a)--(c) of Fig.~\ref{fig:tpd_cod}, showing only the $m/z$ = 45 ([HCS]$^+$) and 46 ([H$_2$CS]$^+$) signals. Arrows mark the weak, coincident features between approximately 70 and 95~K, attributed to desorption of monomeric H$_2$CS.}
    \label{fig:h2cs-monomer-desorption}
\end{figure}

Figure~\ref{fig:h2cs-monomer-desorption} shows an enlarged view of the TPD-QMS data. A weak, coincident rise in the $m/z$ = 45 and 46 signals is observed between approximately 70 and 95~K, consistent with the desorption of a small fraction of monomeric H$_2$CS embedded in the ice.

\end{appendix}

\end{document}